\documentclass[lettersize,journal]{IEEEtran}
\usepackage{amsmath,amsfonts}
\usepackage{algorithmic}
\usepackage{algorithm}
\usepackage{array}
\usepackage[caption=false,font=normalsize,labelfont=sf,textfont=sf]
{subfig}
\usepackage{textcomp}
\usepackage{stfloats}
\usepackage{url}
\usepackage{verbatim}
\usepackage{graphicx}
\usepackage{cite}
\usepackage[colorlinks=true, linkcolor=blue, citecolor=blue, urlcolor=blue, breaklinks=true]{hyperref}
\usepackage{orcidlink}
\usepackage{multirow}
\usepackage{soul}
\usepackage{enumitem}
\usepackage{diagbox} 
\usepackage{makecell} 

\graphicspath{{pictures/}}

\usepackage{listings}
\usepackage{xcolor}
\definecolor{customcolor}{HTML}{54278f}
\definecolor{foo}{HTML}{EEEEFF}
\definecolor{mygreen}{HTML}{1b9e77}

\usepackage{fancyhdr}

\begin{document}

\title{VIGMA: An Open-Access Framework for\\Visual Gait and Motion Analytics}

\author{Kazi Shahrukh Omar\orcidlink{0000-0003-3182-6553}, Shuaijie Wang, Ridhuparan Kungumaraju, Tanvi Bhatt, Fabio Miranda\orcidlink{0000-0001-8612-5805}
\thanks{Kazi Shahrukh Omar, Ridhuparan Kungumaraju, and Fabio Miranda are with the Department of Computer Science, and Shuaijie Wang and Tanvi Bhatt are with the Department of Applied Health Sciences, University of Illinois Chicago, IL 60607, USA (emails: komar3@uic.edu; sjwang4@uic.edu; rkungu2@uic.edu; tbhatt6@uic.edu; fabiom@uic.edu;) 
}
}

\markboth{Accepted for publication in IEEE Transactions on Visualization and Computer Graphics}%
{Shell \MakeLowercase{\textit{et al.}}: A Sample Article Using IEEEtran.cls for IEEE Journals}


\maketitle

\begin{abstract}

Gait disorders are commonly observed in older adults, who frequently experience various issues related to walking. Additionally, researchers and clinicians extensively investigate mobility related to gait in typically and atypically developing children, athletes, and individuals with orthopedic and neurological disorders. Effective gait analysis enables the understanding of the causal mechanisms of mobility and balance control of patients, the development of tailored treatment plans to improve mobility, the reduction of fall risk, and the tracking of rehabilitation progress. However, analyzing gait data is a complex task due to the multivariate nature of the data, the large volume of information to be interpreted, and the technical skills required. Existing tools for gait analysis are often limited to specific patient groups (e.g., cerebral palsy), only handle a specific subset of tasks in the entire workflow, and are not openly accessible. To address these shortcomings, we conducted a requirements assessment with gait practitioners (e.g., researchers, clinicians) via surveys and identified key components of the workflow, including (1) data processing and (2) data analysis and visualization. Based on the findings, we designed VIGMA, an open-access visual analytics framework integrated with computational notebooks and a Python library, to meet the identified requirements. Notably, the framework supports analytical capabilities for assessing disease progression and for comparing multiple patient groups. We validated the framework through usage scenarios with experts specializing in gait and mobility rehabilitation. VIGMA is available at \href{https://github.com/komar41/vigma}{github.com/komar41/VIGMA}. 

\end{abstract}



\begin{figure*}[t]
  \centering
  \includegraphics[width=1\linewidth]{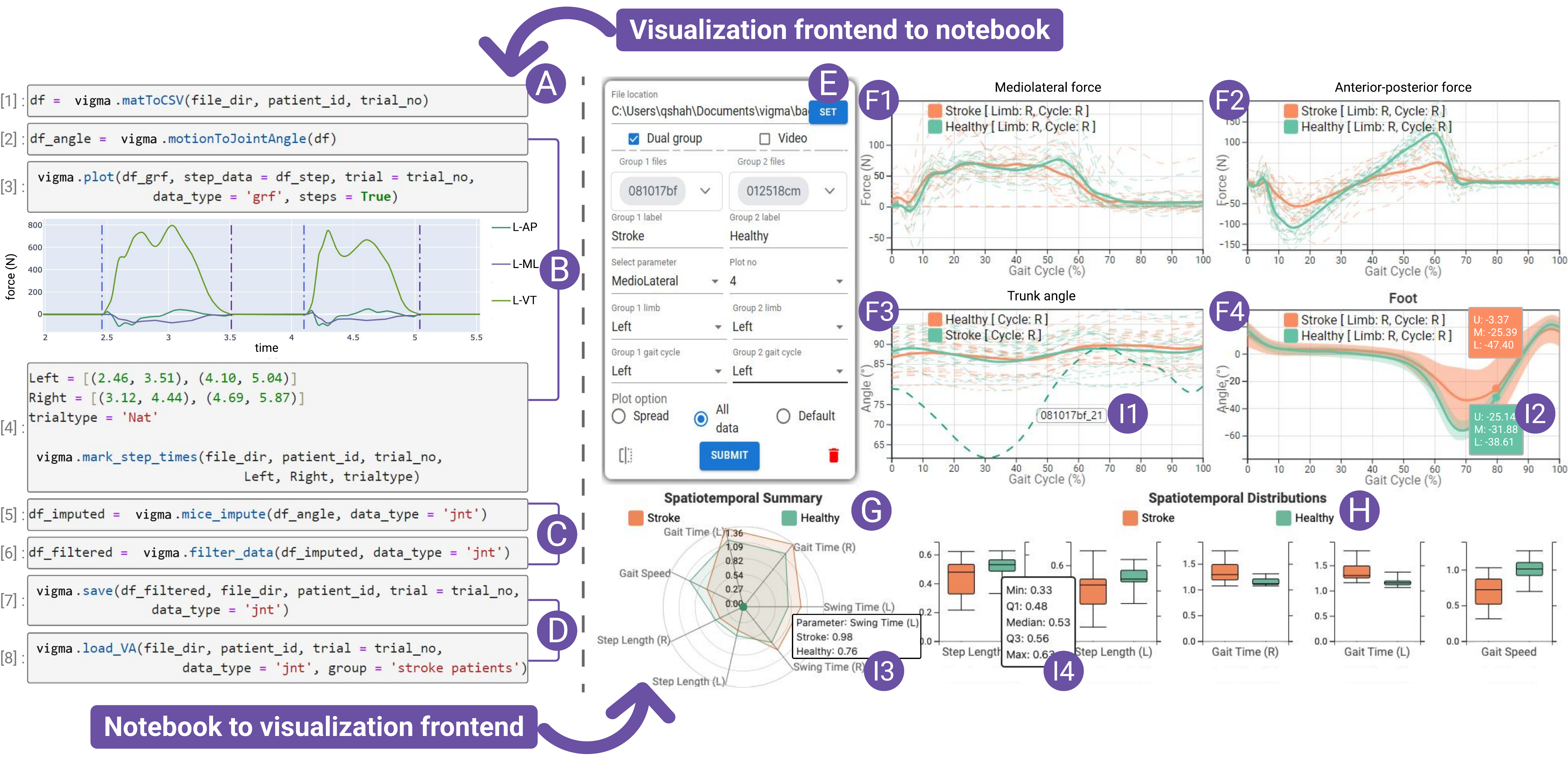}
  \caption{
  \textbf{VIGMA}'s components. Using Jupyter Notebook and VIGMA's library, users begin by \lower0.2em\hbox{\includegraphics[width=1em]{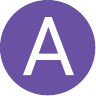}} harmonizing collected data to standard CSV format. Afterwards, they perform essential \lower0.2em\hbox{\includegraphics[width=1em]{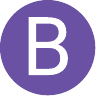}} feature extraction steps (e.g., extracting joint angles from motion data, extracting step times from ground reaction forces). Next, they prepare the data by performing tasks like \lower0.2em\hbox{\includegraphics[width=1em]{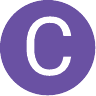}} imputing missing values and filtering out noise. Finally, \lower0.2em\hbox{\includegraphics[width=1em]{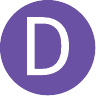}} users save the processed data and upload it to the visual analytics system, categorized by patient group. The visual analytics system consists of \lower0.2em\hbox{\includegraphics[width=1em]{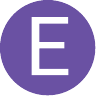}} a control panel that lets users select ensemble data files and select chart configurations. The system is composed of four time-series ensemble views \lower0.2em\hbox{\includegraphics[width=1em]{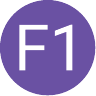}} - \lower0.2em\hbox{\includegraphics[width=1em]{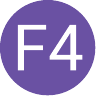}} to support the analysis of time-series gait data (e.g., joint angles). Additionally, it features \lower0.2em\hbox{\includegraphics[width=1em]{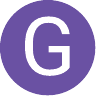}} a spatiotemporal summary view and \lower0.2em\hbox{\includegraphics[width=1em]{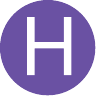}} a spatiotemporal distribution view to support single-valued gait parameter analysis. Users can choose to alternate between the spatiotemporal views (\lower0.2em\hbox{\includegraphics[width=1em]{icons/g.pdf}}, \lower0.2em\hbox{\includegraphics[width=1em]{icons/h.pdf}}) and the video exploration view (Fig.~\ref{fig:video-exploration-view}) using the checkbox from the control panel \lower0.2em\hbox{\includegraphics[width=1em]{icons/e.pdf}} and get access to raw trial videos. \lower0.2em\hbox{\includegraphics[width=1em]{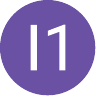}} to \lower0.2em\hbox{\includegraphics[width=1em]{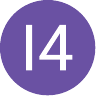}} highlight interactive details-on-demand features of the system. Users can identify anomalies (e.g., \lower0.2em\hbox{\includegraphics[width=1em]{icons/i1.pdf}}), trace errors to patients' trial videos, and return to the computational notebook to correct the data.}
  \label{fig:teaser}
\end{figure*}

\section{Introduction}
Gait disorders are prevalent in older adults, with studies revealing that 82\% of people aged 85 years or older suffer from some sort of gait anomaly~\cite{bloem1992investigation, sudarsky2001gait}. Conditions such as Parkinson's~\cite{nieuwboer1998frequency} or stroke~\cite{eng2002reliability} significantly impact patients' gait, causing notable deviations from typical motor movements such as shuffling or asymmetric gait. Additionally, gait impairments are increasingly observed in children with cerebral palsy~\cite{himmelmann2010changing} and spina bifida~\cite{mitchell2004spina}. The analysis of gait presents a complex challenge due to its variability within and across different patient groups, coupled with the diverse types of data collected for each individual. Consequently, tools that straddle the entire gait data lifecycle (processing, analyzing, and visualizing) can play a pivotal role in understanding the biomechanics of a patient, monitoring disease progression, and implementing appropriate rehabilitation strategies.

Gait data is collected via a diverse combination of hardware (e.g., infrared (IR) cameras, force plates) and software tools (e.g., Visual3D~\cite{visual3d}, Vicon Nexus~\cite{vicon}) that generate a wide variety of time-series data, including kinetics (ground reaction forces, joint moment), kinematics (motion, joint angles), and electromyography, as well as single-valued spatiotemporal parameters (e.g., step length, swing time). Each of these data types plays a crucial role in various tasks associated with the analysis of patients' gait. For instance, kinetic measures and joint moments can provide insights into several musculoskeletal diseases~\cite{karatsidis2016estimation}, whereas kinematic measures are particularly useful for assessing fall risk in elderly individuals~\cite{hamacher2011kinematic}.
Gait data also includes raw trial videos, which, when analyzed alongside other time-series data, provide additional context to interpret anomalies or validate findings.
This diversity of data underscores the need for comprehensive tools that encompass all of these data types to enable a more detailed analysis of patients' gait.

Therefore, analyzing gait data constitutes a multifaceted challenge, as the workflow involves several data processing, and data analysis and visualization tasks. Data processing tasks involve format harmonization, imputing missing values, filtering noise, whereas analysis and visualization tasks include analyzing statistical measures, comparing patient groups, finding anomalies, and analyzing disease progression. Both sets of tasks are vital in the overall workflow of practitioners. However, existing tools and techniques predominantly support only a subset of tasks and lack important analysis and visualization capabilities, such as comparing patient groups and analyzing disease progression. Furthermore, the lack of open access to tools utilized by practitioners in different labs hinders collaborative efforts between groups.

To tackle these challenges, this paper presents the open-access \ul{Vi}sual \ul{G}ait and \ul{M}otion \ul{A}nalytics (VIGMA) framework. VIGMA was designed in close collaboration with domain experts and is the result of a survey with over 20 gait practitioners, including clinicians, researchers, administrators, data analysts, faculty members, and students.
The survey allowed us to surface detailed workflows for analyzing gait data that involved a wide variety of tasks, including data processing, and data analysis and visualization. 
In response to the wide needs of gait practitioners, VIGMA is designed as both (a) a library for use in computational notebooks, enabling data processing tasks, and (b) a visual analytics system for data analysis and visualization.
The visual analytics system supports the visualization of multivariate time-series ensemble gait data, along with the analysis of video data and spatiotemporal parameters with tailored interaction techniques to support tasks such as group comparison and tracking disease progression.
Fig.~\ref{fig:teaser} illustrates the VIGMA framework interface, highlighting its individual components and their functionalities.
This paper also presents an evaluation of VIGMA that includes three usage scenarios created in collaboration with domain experts, as well as experts' quantitative and qualitative feedback on the framework.
Our contributions can be summarized as follows: 

\begin{itemize}[noitemsep, nolistsep]
    \item We present a comprehensive survey yielding requirements for gait data processing, analysis, and visualization.
    \item We contribute a Python library that provides functionalities to satisfy gait data processing requirements.
    \item We contribute a novel visual analytics system that supports the data analysis and visualization requirements.
    \item We present three usage scenarios created in collaboration with domain experts to evaluate the efficacy of the system.
\end{itemize}

This paper is organized as follows: Section~\ref{sec:background} presents background on gait data and analysis; Section~\ref{sec:related} reviews related work; Section~\ref{sec:surface-workflow-req} presents the result of our survey and requirements; Section~\ref{sec:vigma} details the VIGMA framework; Section~\ref{sec:evaluation} presents the evaluation; Section~\ref{sec:conclusion} concludes the paper.

\section{Background}
\label{sec:background}

Gait, or walking, is a vital social activity controlled by an intricate combination of brain activities and limb movements. It is commonly studied in the elderly, children, and athletes due to the unique challenges and needs these groups present. Practitioners collect gait data, categorized into videos, kinematics, kinetics, electromyography, and spatiotemporal parameters. Monthly intervals are typical for data collection, varying based on disease and severity~\cite{vervoort2016progression, mohan2021assessment}.

\begin{figure}[t]
  \centering
  \includegraphics[width=1\linewidth]{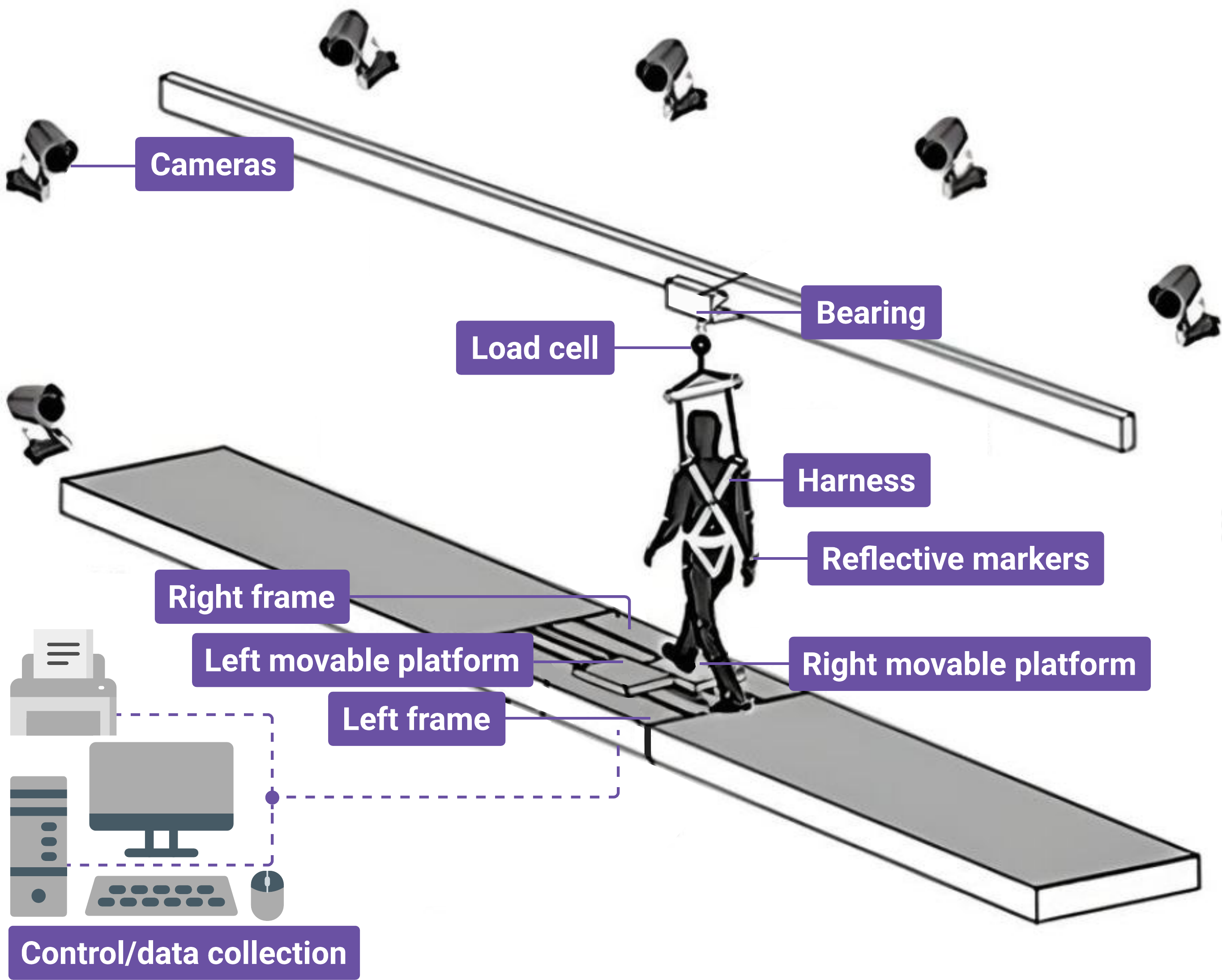}
  \caption{Data collection process~\cite{yang2013generalization, pai2014perturbation, wang2019can,wang2022near}. A 7-meter walkway with embedded low-friction movable platforms is used to collect slip and walking data. The platforms move along low-friction aluminum tracks beneath the surface and rest on four force plates. The force plates record ground reaction forces, which also trigger the platforms to create the slipping effect. The platforms lock firmly during regular walking and unlock electromechanically without the participants’ awareness in the slip trial. Participants are protected by a safety harness connected to a low-friction trolley-and-beam system above the walkway. Kinematic data from all trials is recorded by an eight-camera motion capture system, synchronized with the force plates and load cell data.}
  \label{fig:background}
\end{figure}

Practitioners tailor their collection of data types to the specific needs of different patient groups. For instance, post-stroke patients exhibit a spectrum of characteristics, such as decreased walking speed and shorter stride length~\cite{jacquelin2010gait}. They also encounter heightened fall risks due to impaired balance control and muscle weaknesses~\cite{dobkin2005rehabilitation}. In the case of cerebral palsy, patients showcase distinctive walking patterns (e.g., jump knee gait, equinus~\cite{sutherland1993common}), which can be identified by analyzing kinematic gait data~\cite{chang2010role}. Parkinson's patients, on the other hand, face noticeable gait deterioration in kinematic (e.g., ankle, knee angle) and spatiotemporal parameters (e.g., reduced step length and arm swing)~\cite{pistacchi2017gait}.

Following such diversity of needs, data is collected in different scenarios, including regular and perturbed (slip/trip) walking. Fig. \ref{fig:background} illustrates a typical data collection process~\cite{yang2013generalization, pai2014perturbation, wang2019can,wang2022near}. It involves a walkway with embedded movable platforms, which rest on force plates to record kinetic data. These platforms unlock once participants step on them during slip trials, and participants are protected by a safety harness during the trials. An infrared camera-based motion capture system records video and kinematic data by tracking markers placed on the body. Electromyography (EMG) data is collected using electrodes placed on the skin over the muscles of interest, which detect electrical activity during walking. Spatiotemporal data, which includes spatial parameters (e.g., step length) and temporal parameters (e.g., cadence), is often derived from kinematic or kinetic data. All these different data types are synchronized to provide a comprehensive analysis of gait. 

The collected gait data types fall into the categories of time-series or single-valued, with spatiotemporal parameters belonging to the latter. To process and analyze these two different types of data, different software systems are used, facilitating processing, analysis, and visualization of the collected data. 
In terms of analysis, practitioners typically focus on tasks such as: (1) comparing patients' gait characteristics with those of healthy control groups or similar patients to identify deviations and abnormalities~\cite{jonsdottir2009functional, boudarham2013variations, iosa2016stability}, (2) assessing gait changes over time to monitor disease progression and evaluate the effectiveness of rehabilitation strategies~\cite{beyaert2015gait, moreau2016effectiveness, marin2020my}, (3) statistical analysis~\cite{schwartz2004measurement, wu2009statistical}, and (4) finding anomalies~\cite{wu2009statistical, sangeux2015simple}.

\section{Related Work}
\label{sec:related}

In this section, we review related work across four categories. First, we discuss the existing works on visualization for gait;
next, we explore literature for visualization for time-series data; then, we investigate open-access and commercial tools that support gait data processing, analysis, and visualization workflows;
and finally, we review existing research on computational notebooks to support data processing tasks and how they are combined with visualization interfaces.



\noindent \textbf{Visualization for gait data.}
Gait data can be grouped into (1) time-series~\cite{zakaria2021markerless} and (2) single-valued parameters~\cite{seals2022they}. Spatiotemporal data are considered single-valued parameters, while kinetic, kinematic, and electromyography data are classified as time-series. Existing approaches commonly use line charts~\cite{anwary2020gait, zakaria2021markerless, boumrah2022real, gonzalez2022crouch} for time-series gait data, radar charts~\cite{mukaino2018clinical,seals2022they} for single-valued parameters, and box plots~\cite{anwary2020gait, seals2022they} or parallel coordinate plots~\cite{cantu2023parallel} for the distribution of single-valued parameters. However, these approaches often lack the interactivity and human-in-the-loop capabilities provided by traditional visual analytics tools. In contrast, tools like MotionExplorer~\cite{bernard2013motionexplorer}, MotionFlow~\cite{jang2015motionflow}, and Chen et al.~\cite{chen2020visual} support interactivity to provide insight into patient’s gait and help researchers cluster similar human poses. However, these tools do not provide any insight into quantitative gait measures (e.g., kinetic, kinematic, spatiotemporal) that practitioners use for the analysis and treatment of patients.

Existing literature introduced several novel visual analytics tools for quantitative gait analysis. The NE-Motion visual analytics tool~\cite{contreras2021ne} is designed to compare the gait of healthy individuals with that of stroke patients experiencing upper limb impairments. KAVAGait~\cite{wagner2018kavagait} supports the analysis of kinematic (e.g., ground reaction forces) and spatiotemporal gait in various patient groups. GaitViewer~\cite{agibetov2016gaitviewer} proposes a structured approach to storing numerical gait data, facilitating collaboration between research labs, and providing an interactive visual interface for spatiotemporal gait parameters. gaitXplorer~\cite{rind2022trustworthy} is a visual analytics tool supporting kinematic gait analysis of cerebral palsy patients. It supports classifying patients into different groups (e.g., jump knee, equinus) and provides insights into the specific data regions that contributed to the classification.
Although useful, these tools offer very limited data processing capabilities.
None provides comprehensive multivariate feature support, and only KAVAGait and gaitXplorer support hierarchical data management. Also, only GaitViewer supports feature extraction, but is limited to a few spatiotemporal parameters. Furthermore, these tools lack important data analysis and visualization capabilities. Specifically, KAVAGait and NE-Motion do not support disease progression analysis; GaitViewer lacks patient group comparison, while gaitXplorer lacks both.
Lastly, none of these tools are openly available.

\noindent \textbf{Visualization for time-series data.} Time-series gait is multivariate, including diverse attributes such as joint angles, forces, and electromyography signals, analyzed across different limbs and normalized over gait cycles. These attributes are often grouped into ensembles for analyzing similar patient groups (e.g., stroke/healthy) collectively or comparing between two groups. Numerous studies have delved into the visual analysis of multivariate time-series data~\cite{fujiwara2020visual, shi2023supporting, liu2024relation, shirato2023exploring, chen2024fmlens, liu2022mtv, stopar2018streamstory}, as well as time-series ensemble data~\cite{ye2022visualizing, deng2023visualizing, yu2022pseudo}. In gait studies, line charts are the most widely used visualization technique for time-series data~\cite{wagner2018kavagait, rind2022trustworthy, visual3d}, making them a familiar choice for gait practitioners in research labs or clinical settings. In this work, we explore design variations and interactions with line charts to effectively represent multivariate and time-series ensemble gait data, supporting the analysis of individual patient’s gait as well as tasks such as disease progression and group comparisons.

Time-series gait analysis also includes videos that offer raw trial information to better understand the data. For instance, videos allow users to identify the reasons behind anomalies or outliers, such as issues with sensors or a misstep on the platform. Similarly, for disease progression analysis, raw trial videos allow users to observe and validate changes in joint angles or ground reaction forces. While many studies have explored visual analytics approaches for analyzing video data~\cite{wu2023liveretro, stein2017bring, zeng2019emoco, chen2021augmenting, he2023videopro, wong2023anchorage}, existing literature and tools for gait~\cite{wagner2018kavagait, rind2022trustworthy, visual3d, dixon2017biomechzoo} often overlook the integration of time-series video information. To address this gap, we explore visual analysis and interaction techniques that allow users to correlate raw video data with multivariate and time-series ensemble information.

\begin{table*}[t]
\centering
\caption{Overview of how open-access and commercial tools, along with tools from existing literature, address different data processing, analysis, and visualization features. Open source refers to software that is freely accessible and allows code inspection. The notations used in the table signify the following: $\surd$ - feature is fully supported, $\triangle$ - feature is partially supported, and empty cells indicate that the respective tool does not support the feature.}

\begin{tabular}{|l|c|c|c|c|c|c|c|}
\hline
\textbf{Features}
& \makecell{\textbf{Kinovea}\\\cite{kinovea}}
& \makecell{\textbf{bioMechZoo}\\\cite{dixon2017biomechzoo}}
& \makecell{\textbf{OpenCap}\\\cite{uhlrich2023opencap}}
& \makecell{\textbf{GaitLab}\\\cite{kidzinski2020deep}}
& \makecell{\textbf{Vicon Nexus}\\\cite{vicon}}
& \makecell{\textbf{TEMPLO}\\\cite{templo}}
& \makecell{\textbf{BioSensics}\\\cite{biosensics}}\\\hline

\multicolumn{1}{|l|}{Format harmonization}          & & $\triangle$ & & & & &  \\ \hline
\multicolumn{1}{|l|}{Feature extraction}            & $\triangle$ & $\surd$ & $\surd$ & $\triangle$ & $\surd$ & $\triangle$ & $\triangle$  \\ \hline
\multicolumn{1}{|l|}{Missing value imputation}      & & $\surd$ & & & $\surd$ & &  \\ \hline
\multicolumn{1}{|l|}{Noise filtering}               & & $\surd$ & & & $\surd$ & &  \\ \hline
\multicolumn{1}{|l|}{Gait cycle normalization}      & & $\surd$ & & & & $\surd$ &  \\ \hline
\multicolumn{1}{|l|}{Hierarchical data management}  & & & & & $\surd$ & &  \\ \hline
\multicolumn{1}{|l|}{Multivariate feature support}  & $\triangle$ & $\surd$ & $\surd$ & $\triangle$ & $\surd$ & $\surd$ & $\triangle$  \\ \hline
\multicolumn{1}{|l|}{Group comparison}              & & $\triangle$ & & & & &  \\ \hline
\multicolumn{1}{|l|}{Disease progression}           & & $\triangle$ & & & & &  \\ \hline
\multicolumn{1}{|l|}{Statistical analysis}          & & $\triangle$ & & & & &  \\ \hline
\multicolumn{1}{|l|}{Anomaly analysis}              & & $\triangle$ & & & & &  \\ \hline
\multicolumn{1}{|l|}{Access to raw video data}      & $\surd$ & & $\surd$ & & & $\surd$ &  \\ \hline
\multicolumn{1}{|l|}{Interactive analysis}          & $\triangle$ & $\triangle$ & & & & $\triangle$ &  \\ \hline
\multicolumn{1}{|l|}{Open source}                   & $\surd$ & & $\surd$ & $\surd$ & & &  \\ \hline
\end{tabular}

\vspace{0.2cm} 

\begin{tabular}{|l|c|c|c|c|c|c|c|}
\hline
\textbf{Features}
& \makecell{\textbf{GAITRite}\\\cite{bilney2003concurrent}}
& \makecell{\textbf{Visual3D}\\\cite{visual3d}}
& \makecell{\textbf{NE-Motion}\\\cite{contreras2021ne}}
& \makecell{\textbf{KavaGait}\\\cite{wagner2018kavagait}}
& \makecell{\textbf{GaitViewer}\\\cite{rind2022trustworthy}}
& \makecell{\textbf{gaitXplorer}\\\cite{agibetov2016gaitviewer}}
& \makecell{\textbf{VIGMA}\\(This work)}\\\hline

\multicolumn{1}{|l|}{Format harmonization}          & & $\triangle$ & & & & & $\surd$  \\ \hline
\multicolumn{1}{|l|}{Feature extraction}            & $\triangle$ & $\surd$ & & & $\triangle$ & & $\surd$  \\ \hline
\multicolumn{1}{|l|}{Missing value imputation}      & & & & & & & $\surd$  \\ \hline
\multicolumn{1}{|l|}{Noise filtering}               & & $\surd$ & & & & & $\surd$  \\ \hline
\multicolumn{1}{|l|}{Gait cycle normalization}      & $\triangle$ & $\surd$ & & & & & $\surd$  \\ \hline
\multicolumn{1}{|l|}{Hierarchical data management}  & & $\surd$ & & $\surd$ & & $\surd$ & $\surd$  \\ \hline
\multicolumn{1}{|l|}{Multivariate feature support}  & & $\surd$ & & $\triangle$ & & & $\surd$  \\ \hline
\multicolumn{1}{|l|}{Group comparison}              & & $\triangle$ & $\triangle$ & $\surd$ & & & $\surd$  \\ \hline
\multicolumn{1}{|l|}{Disease progression}           & $\triangle$ & $\triangle$ & & & $\triangle$ & & $\surd$  \\ \hline
\multicolumn{1}{|l|}{Statistical analysis}          & & $\surd$ & $\triangle$ & $\surd$ & & $\triangle$ & $\surd$  \\ \hline
\multicolumn{1}{|l|}{Anomaly analysis}              & & $\surd$ & $\triangle$ & $\surd$ & $\triangle$ & $\surd$ & $\surd$  \\ \hline
\multicolumn{1}{|l|}{Access to raw video data}      & $\surd$ & & & & & & $\surd$  \\ \hline
\multicolumn{1}{|l|}{Interactive analysis}          & & $\triangle$ & $\triangle$ & $\surd$ & $\triangle$ & $\surd$ & $\surd$  \\ \hline
\multicolumn{1}{|l|}{Open source}                   & & & & & & & $\surd$  \\ \hline

\end{tabular}

\label{tab:tools-and-techniques}
\end{table*}

\noindent \textbf{Open-access and commercial tools.}
Kinovea~\cite{kinovea}, biomechZoo~\cite{dixon2017biomechzoo}, OpenCap~\cite{uhlrich2023opencap}, and GaitLab~\cite{kidzinski2020deep} are a few open-access tools tailored for gait data. Kinovea supports sports analysis by using videos to capture joint angles and extract a few spatiotemporal parameters (gait speed and stride length), but does not support other data processing tasks or the analysis of more than one trial at a time. OpenCap and GaitLab are tools primarily designed for markerless gait data collection using smartphone camera videos. GaitLab is limited to kinematic gait, whereas OpenCap also estimates force data. However, both tools lack capabilities for data processing tasks such as structured access to patient and trial data, loading external formats (e.g., C3D, TRC), or further processing (e.g., noise filtering) of collected data. Furthermore, like Kinovea, they also lack the support for analyzing more than one trial at a time. Among these three tools, Kinovea and OpenCap provide access to raw video data. biomechZoo~\cite{dixon2017biomechzoo}, on the other hand, is an open-access toolbox that supports processing, analyzing, and visualizing different types of gait data, including kinetic, kinematic, spatiotemporal, and electromyography. The software consists of MATLAB scripts that perform data processing and two graphical interfaces that support the analysis and visualization of the data. However, MATLAB itself is not an open-source or free-access software. Additionally, biomechZoo has complex documentation with numerous functions and generates non-interactive charts, which limits its ability to support tasks related to disease progression and group comparisons.

Commercial gait software~\cite{biosensics, bilney2003concurrent, vicon, templo} are equipped with proprietary hardware to collect data and are more focused on accurate data collection than data processing, analysis, and visualization tasks. For instance, BioSensics~\cite{biosensics} offers real-time gait data collection and monitoring tools like LEGSys and BALANSys for assessing gait steadiness and balance, respectively, but provides little to no data processing or data analysis and visualization support. GAITRite~\cite{bilney2003concurrent}, a data collection tool for spatiotemporal parameters, similarly offers minimal support for data processing, analysis, and visualization. Vicon Nexus~\cite{vicon} is another commercial data collection tool for time-series gait data that supports data processing steps such as feature extraction (e.g., spatiotemporal parameters from ground reaction forces), imputing missing values, and filtering noise, but provides no analysis or visualization support. Templo~\cite{templo} is a commercial markerless data collection tool that offers single-trial analysis support by visualizing kinematic, kinetic, and electromyography data alongside the corresponding trial video. While Templo supports certain tasks, such as feature extraction and gait cycle normalization, and provides access to raw video data, it is limited to analyzing only one trial at a time.


Among commercial software, Visual3D~\cite{visual3d} closely aligns with the proposed framework, offering data processing, analysis, and visualization support for multivariate gait data. However, it supports only a limited range of data formats (C3D and ASCII formats) and does not accommodate TRC or MAT formats. Additionally, it lacks missing value imputation capabilities, provides limited analysis support for time-series gait data with non-interactive charts, and does not support analysis of single-valued gait parameters. The software has lengthy documentation and complex code syntax for processing and analyzing files, as well as a complicated user interface. The proposed system, VIGMA, is an open-access tool that addresses these gaps in existing solutions.

We summarize the commercial and open-access tools, including tools from literature~\cite{contreras2021ne, wagner2018kavagait, agibetov2016gaitviewer, rind2022trustworthy}, in terms of
data processing, analysis, and visualization features in Table~\ref{tab:tools-and-techniques}. 
In Section~\ref{subsec:state-of-the-art}, we present a detailed comparison between VIGMA and a selected subset of these tools that we consider to be the state of the art.


\noindent \textbf{Computational notebooks.} Computational notebooks, such as Jupyter, have gained widespread popularity for exploratory data analysis and visualization, with users embracing this approach across various levels of coding expertise. However, static visualizations, dependency between cell execution order, and scrolling through too many cells make analysis tedious for practitioners. Recent works~\cite{lee2021lux,tritsarolis2021st_visions, chen2022pi2,li2023edassistant, epperson2023dead} have attempted to address these issues by leveraging code in computational notebooks while visualizing data separately in interactive dashboards, linking them together to combine the strengths of both. Similarly, in the domain of machine learning interpretability, many recent works~\cite{palmeiro2022data+, wang2022interpretability, xenopoulos2022calibrate, guo2023causalvis, guo2021vaine} have integrated notebooks with interactive visualization dashboards to enhance the interpretation, calibration, and validation of machine learning models.

Working closely with gait experts and surveying diverse gait practitioners across research labs and clinics, we found that they rely on computational notebooks, such as Live Scripts in MATLAB or Jupyter in Python, for data processing tasks. To that end, we adopt a similar approach to these existing studies, handling data processing tasks in computational notebooks and integrating them with the proposed visual analytics system that supports analysis and visualization tasks. The distinction between this work and these studies lies in the fundamental consideration of the problem space. The proposed system is specifically designed for the gait workflows, with unique data processing, analysis, and visualization requirements.
\section{Surfacing Workflows and Requirements}
\label{sec:surface-workflow-req}

Following design methodologies from prior works~\cite{li2023causality, teng2023vispur, jiang2023healthprism}, we designed a semi-structured survey in collaboration with two researchers specializing in gait studies (co-authors of this paper) to surface the overall workflow and requirements of practitioners.
The survey included a set of questions aimed at collecting insights into how practitioners process gait data, as well as their analysis and visualization needs. 
%
The survey was conducted using an online form supplemented by follow-up open-ended questions. Approval for the study was obtained from UIC's Institutional Review Board (\#2016-0933).
The primary questions asked of participants were:

\begin{itemize}[noitemsep, nolistsep]
    \item What target population do you collect data from?
    \item What is the primary goal of your data collection?
    \item What type of gait data do you typically work with?
    \item What are the formats of data you collect?
    \item What data processing tasks do you perform?
    \item What data analysis tasks do you perform?
\end{itemize}

In addition, we also asked participants to evaluate current visualization designs used for interpreting gait data.
We had a total of 23 participants for the survey, with diverse job titles and from different research labs and clinics.
The majority of survey responses, comprising 15 out of 22 (65.2\%), were from researchers. Among these 15 respondents, six also identified as clinicians and six as students. Completing the survey took participants approximately 25 to 30 minutes. 
The survey questionnaire, participant responses, and the corresponding requirements extracted from these responses are summarized in Table I of the appendix. The eleven visualizations presented to the participants and the evaluations from the participants are presented in Fig. 1 and Table II of the appendix, respectively.

After the survey was completed, we inspected the results to surface a set of requirements for the framework. This process was conducted through meetings between the authors of this paper, which included both visualization researchers and domain experts.
We derived three data processing requirements (\textbf{R1} - \textbf{R3}) and six data analysis \& visualization requirements (\textbf{R4} - \textbf{R9}). 
To address the needs of a broad user group comprising gait practitioners from different research labs and clinics, we sought to identify their visual encoding preferences. Building on existing studies~\cite{yang2014understand, quadri2024you, zhu2024compositingvis} that emphasize comprehension, usability, and user preferences in visualization design, we presented the participants with eleven visualizations (Fig. 1 in the appendix).
These visualizations were selected after a review of existing literature and gait analysis and visualization tools~\cite{jansen2012similar, serrao2016gait, blau2017quantifying, wagner2018kavagait, troisi2021synthetic, rind2022trustworthy, seals2022they}. We asked participants to evaluate them in terms of usability and understandability. We believe such practices are valuable in design studies of visualization, as supporting a diverse group of domain experts with limited visualization experience requires using visual designs that prioritize usability and comprehension. Furthermore, existing works~\cite{kwon2018retainvis, linhares2022clinicalpath,wentzel2024ditto} highlight that relying on familiar designs rather than novel glyphs is more successful with wider audiences, especially with healthcare experts. Thus, we selected familiar visual designs among experts through a literature review and then refined the selection for the proposed system based on the survey results.

The findings on visual design preferences are outlined in Section~\ref{subsec:vis-encoding-pref}, followed by the data processing, and data analysis and visualization requirements extracted from the survey in Section~\ref{subsec:data-processing-req}.


\subsection{Findings on Visual Design Preferences}
\label{subsec:vis-encoding-pref}

The eleven visualizations presented to the survey participants support displaying time-series gait data (e.g., joint angles, ground reaction forces) as well as single-valued spatiotemporal parameters (e.g., gait speed, swing time). 
%
We asked the participants to evaluate these visualizations in terms of usability and understandability.
Based on the mean user scores, along with the additional comments provided by the participants, we derived the following findings:

\begin{itemize}[noitemsep, nolistsep]
    \item To analyze time-series gait data, existing literature and tools almost solely rely on line plots (Fig. 1-A-F in the appendix). The survey participants were also found to be accustomed to using line plots, and, on average, scored them comparably higher than other types of visualizations.
    \item To analyze the distribution of single-valued parameters such as sociodemographic or spatiotemporal gait parameters, the participants expressed preference for box plots (Fig. 1-I in the appendix) over strip plots (Fig. 1-G, H in the appendix), finding strip plots to be more complex and less intuitive.
    \item For displaying values and making comparisons involving spatiotemporal parameters, participants found the radar plot (Fig. 1-J in the appendix) to be highly useful. This visualization method is particularly beneficial for representing single-valued gait parameters relevant to rehabilitation or disease progression across multiple trials or when comparing different patient groups, such as stroke patients and older adults.
    
\end{itemize}

We leveraged the visualization preferences of the survey to tailor familiar solutions (i.e., line, box, and radar plots) that satisfy the
requirements of the framework. The design of visual encodings and interactions is discussed in detail in Section~\ref{subsec:frontend}.

\subsection{Requirements}
\label{subsec:data-processing-req}

Here, we describe the extracted data processing requirements in \textbf{R1} – \textbf{R3}, followed by the data visualization and analysis requirements in \textbf{R4} – \textbf{R9}.

\noindent \textbf{R1. Data standardization.} Gait data originates from different hardware and software systems like Vicon Nexus~\cite{vicon} and TEMPLO~\cite{templo}, generating diverse formats such as MAT, C3D, TRC, TXT, and CSV, which introduces challenges for collaboration and data sharing. To address these, the framework should support different data formats and harmonization features. Additionally, it should support feature extraction tasks, such as deriving joint angles, step times, spatiotemporal parameters, and fall risk predictions from combinations of motion, ground reaction forces, and joint data. Lastly, the framework should offer data preparation support for imputing missing values, filtering noise, and normalizing data to align gait cycles and ensure consistency across trials.

\noindent \textbf{R2. Data management.} Data is gathered from a diverse range of patient groups, such as stroke patients, older adults, and cerebral palsy patients, across various research labs and clinics. To conduct efficient gait data analysis, practitioners need to have access to two critical pieces of information: (1) patient groups and (2) individual patient trials. This enables effective comparisons between different patient groups or various trials of the same patient. Consequently, the proposed framework should support accessing patient data in this hierarchical manner. Furthermore, for analysis tasks, users often need to refer back to the raw trial videos. Therefore, the framework should support multimodal data access, ensuring on-demand availability of raw trial videos.


\noindent \textbf{R3. Computational notebooks.} Practitioners heavily rely on computational notebooks, such as Live Scripts in MATLAB or Jupyter in Python, for their data processing needs. Notebooks allow them to perform data processing tasks, revisit previous steps, check processed data, identify errors, and re-execute steps to ensure the correctness of the processed data. Additionally, notebooks are easily shareable, fostering collaboration. However, practitioners' programming expertise is often limited. Thus, the framework should support performing data processing tasks (\textbf{R1}, \textbf{R2}) through short one-liner functions within computational notebooks.


\noindent \textbf{R4. Multivariate characteristics.} Gait data is collected in diverse forms, including ground reaction forces, 2D/3D motion, joint angles, electromyography, joint moment, spatiotemporal parameters, and more. These data can be categorized into two primary types: time-series and single-valued parameters. To facilitate effective gait analysis, the framework should possess the capability to analyze and visualize both types of gait characteristics of patients.

\noindent \textbf{R5. Group comparison.} Facilitating the comparison of different groups of gait trials allows for a detailed analysis of key characteristic differences between patient groups. For instance, comparing gait data of stroke patients with that of healthy older adults can reveal significant differences, allowing practitioners to focus on improving specific gait attributes in stroke patients. These types of analyses are crucial in gait studies. Therefore, the framework should support the aggregation of trials and the comparison between ensembles.

\noindent \textbf{R6. Disease progression.} Tracking rehabilitation or disease progression of patients over time is a key aspect of gait studies. For instance, practitioners often monitor key gait characteristics of stroke patients over a period of 6-12 months, adjusting rehabilitation strategies based on improvements or deteriorations in the patient's condition. This approach is similarly applied to injured athletes, cerebral palsy patients, and others. Consequently, the framework should enable seamless analysis of rehabilitation and disease progression of patients.

\noindent \textbf{R7. Statistical measures.} The analysis of statistical measures of gait data is important in acquiring insights into data distribution and facilitating comparisons between different patient groups. In addition, it enables the assessment of how a patient trial deviates from the distribution of a different group or within its respective group. To support these aspects, the framework should enable the analysis of statistical measures, including confidence interval for time-series gait data and mean, median, percentiles, and standard deviation for single-valued parameters.

\noindent \textbf{R8. Anomaly analysis.} Identifying anomalies in gait data is crucial for understanding unusual patient behaviors and ensuring data accuracy. For instance, detecting outliers in the data can help tailor training or clinical interventions to address specific patient needs. Additionally, identifying erroneous data (e.g., missing data, noise) prompts necessary corrective actions to maintain data integrity. The framework should support the detection of outliers and erroneous data, as well as provide corrective measures to fix the errors.

\begin{figure}[t]
  \centering
  \includegraphics[width=1\linewidth]{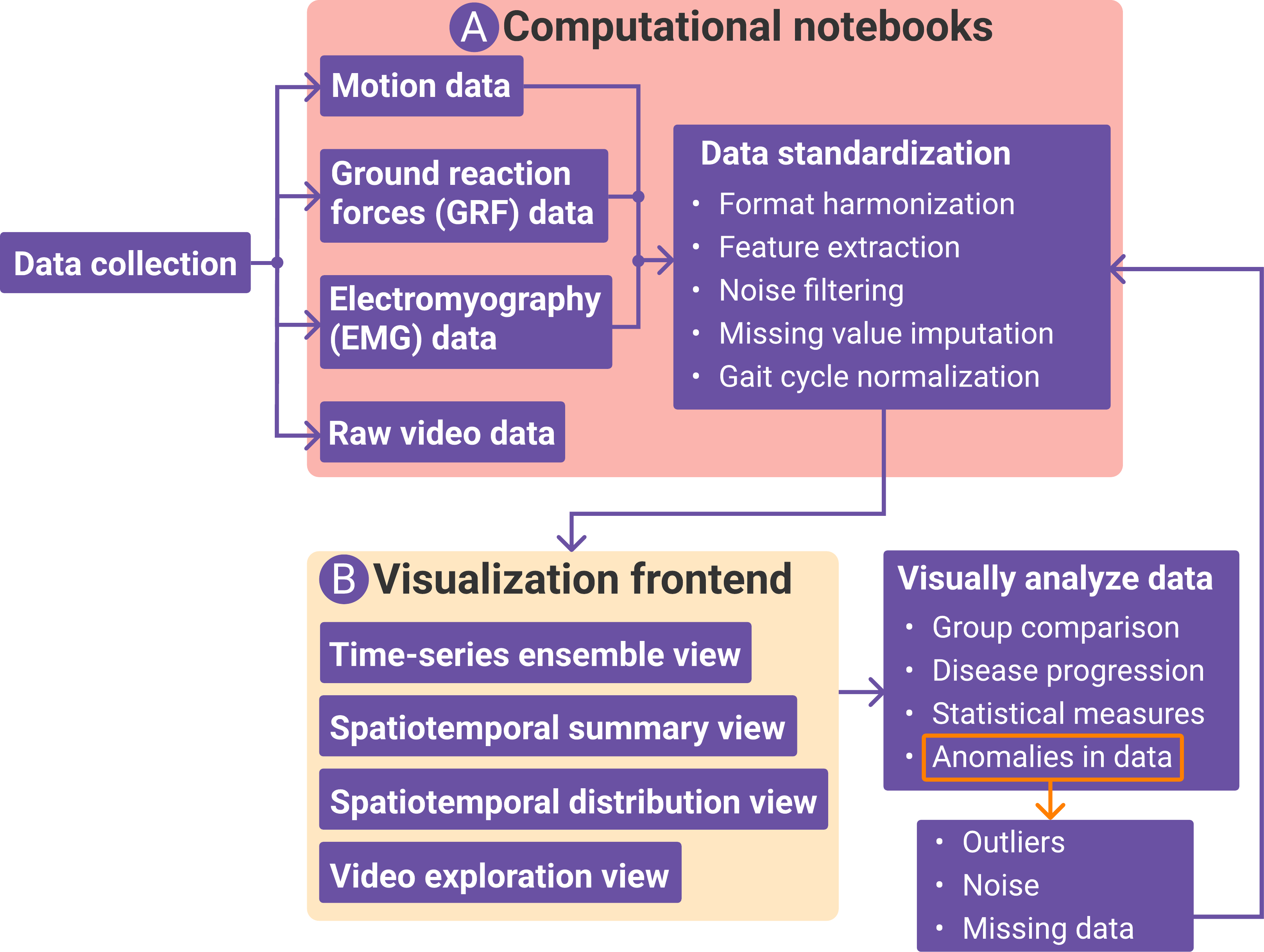}
  \caption{Overview of the workflow. The framework comprises two main components: \lower0.2em\hbox{\includegraphics[width=1em]{icons/a.pdf}} computational notebooks
  and \lower0.2em\hbox{\includegraphics[width=1em]{icons/b.pdf}} visualization frontend. The notebooks facilitate processing gait data (e.g., motion data) by performing standardization steps (e.g., feature extraction, noise filtering) before sending it to the visualization frontend. Users can interact with the frontend to select ensembles of trials and generate charts for time-series and spatiotemporal gait data using time-series ensemble view, spatiotemporal summary view, and spatiotemporal distribution view and have access to raw video data through video exploration view. If anomalies are identified in the data, users can return to the notebooks to correct the data and then visualize the corrected data in the frontend.}
  \label{fig:workflow}
\end{figure}

\noindent \textbf{R9. Interactive analysis.} The framework should support selection, filtering, highlighting, overview-first and details-on-demand, as well as cross-view interactions, to enable human-in-the-loop workflows~\cite{endert2014human} for analysis and visualization tasks (\textbf{R4 – R8}). For multivariate analysis (\textbf{R4}), users should be able to \textbf{select} different gait variables and parameter configurations. For group comparisons (\textbf{R5}) or tracking disease progression (\textbf{R6}), users should be able to \textbf{filter} from the entire dataset, \textbf{selecting} ensembles of trials. To identify and analyze anomalies (\textbf{R8}), users should have the ability to \textbf{highlight} specific trials or single-valued parameters in the visualizations and view \textbf{details} such as patient IDs and trial numbers on demand. Additionally, the framework should provide access to \textbf{detailed} numeric values, such as mean, median, and confidence intervals, alongside the visualizations (\textbf{R7}). Lastly, \textbf{cross-view interactions} between visualizations would enable users to explore relationships among different gait variables, facilitating more in-depth insights into tasks such as group comparisons (\textbf{R5}), disease progression (\textbf{R6}), and anomaly analysis (\textbf{R8}).

\section{The VIGMA Framework}
\label{sec:vigma}

\begin{figure}[t]
\centering
\begin{lstlisting}[caption={VIGMA Python API to perform data processing tasks.}, label=lst:vigma_api]
import vigma
# Format harmonization: MAT to CSV
df = vigma.matToCSV
# Feature extraction: motion to joint angle
df = vigma.motionToJointAngle
# Data preparation: missing value imputation
df = vigma.mice_impute
# Data preparation: filtering noise
df = vigma.filter_data
# Data management: save data
vigma.save
\end{lstlisting}
\end{figure}

VIGMA was designed based on the requirements collected in Section~\ref{sec:surface-workflow-req}. We followed a human-centered design approach, iteratively prototyping and implementing the system while gathering continuous feedback through bi-weekly one-hour in-person meetings with two gait researchers (co-authors of the paper).
%
Their feedback primarily helped us make the Python API less code-intensive and more comprehensible to domain experts, improve the overall system's layout, and select encoding designs and user-friendly interactions for the visualization frontend.

VIGMA has two main components: (1) A Python library that can be accessed in computational notebooks (Fig.~\ref{fig:workflow}-A), and (2) a visualization frontend (Fig.~\ref{fig:workflow}-B). Data processing tasks are carried out using computational notebooks. The processed data are then accessed and visualized using the frontend. 
%
Fig.~\ref{fig:workflow} details the interactions between the two components.
Sections~\ref{subsec:notebooks}
and~\ref{subsec:frontend} detail the data processing and visual design of the framework, respectively, followed by implementation details in Section~\ref{subsec:implementation}.

\subsection{Data Processing}\label{subsec:notebooks}

To streamline gait data processing, our framework contains a Python library that simplifies data tasks into single-line function calls that can be used in computational notebooks (\textbf{R3}). 
%
Example of the VIGMA Python API is illustrated in Listing \ref{lst:vigma_api}.


The API facilitates data standardization (\textbf{R1}) by converting formats such as TRC, MAT, and C3D into structured dataframes, which can then be saved as CSV files. It supports feature extraction of joint angles, step times, spatiotemporal parameters, joint moments, and fall risk predictions derived from motion data, ground reaction forces, and center of mass. Additionally, it offers support for missing value imputation, noise filtering, and normalization of trials by gait cycles.
The API allows (1) saving trial data using a hierarchical file structure that stores it by patient groups and individual patients and (2) loading it into the visualization frontend (\textbf{R2}).
Users can access raw trial videos (\textbf{R2}) alongside the processed data through the visualization frontend. 

\subsection{Visual Design}
\label{subsec:frontend}

\begin{figure}[b]
  \centering
  \includegraphics[width=0.5\linewidth]{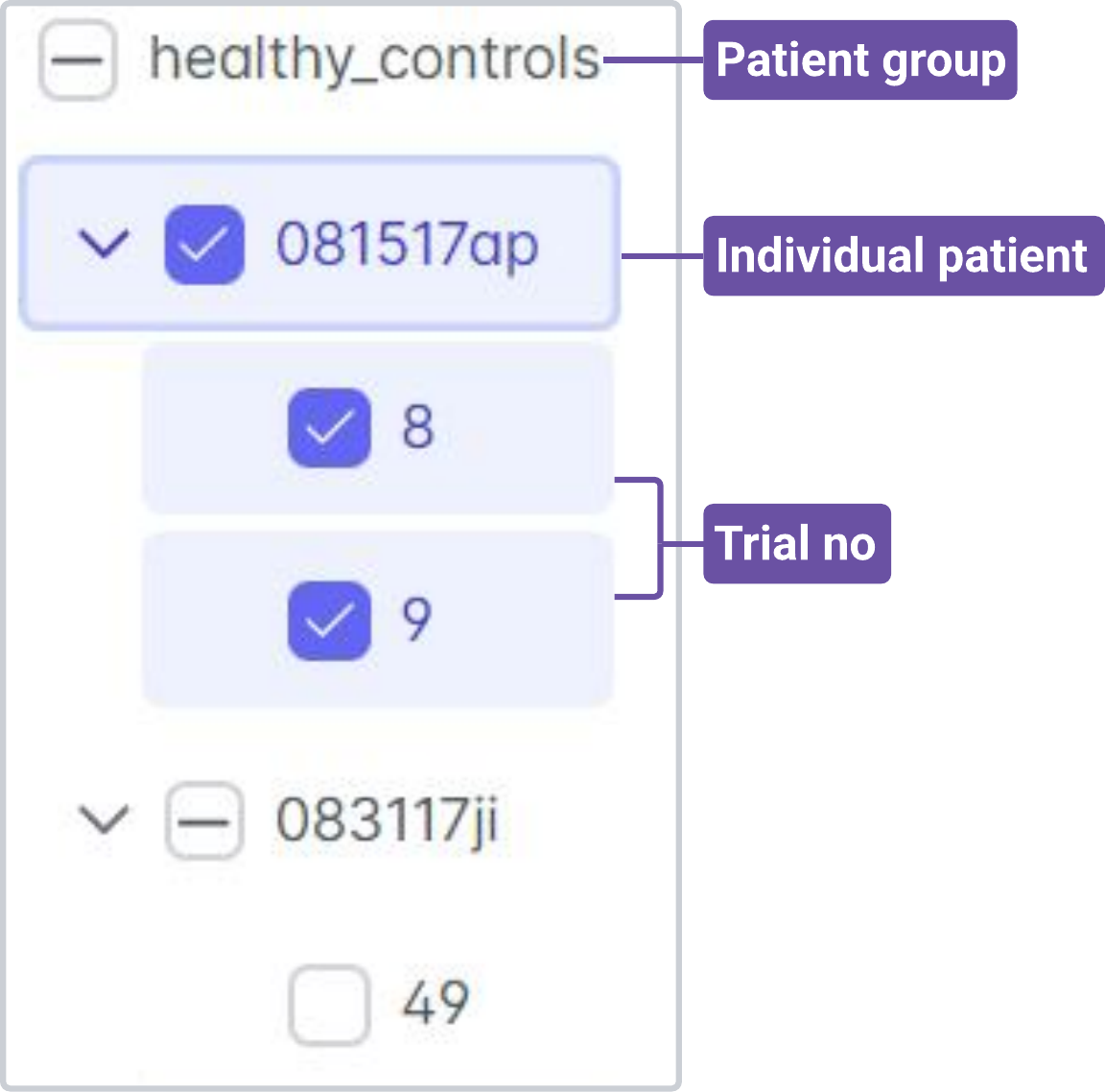}
  \caption{Accessing data through the control panel of the visualization frontend with each trial hierarchically organized by patient ID and patient group.}
  \label{fig:file-structure}
\end{figure}

The frontend of VIGMA is composed 
of three different views: (1) time-series ensemble view, (2) spatiotemporal summary view, and (3) spatiotemporal distribution view. The spatiotemporal summary and distribution views can be alternatively swapped altogether with the video exploration view using a checkbox from the control panel (Fig.~\ref{fig:teaser}-E). The video exploration view provides access to raw trial videos that are linked to the trials from the time-series ensemble views (\textbf{R2}). The frontend also includes a control panel that allows seamless access and filtering of patients' trial data through the hierarchical structure, organized by patient and patient group (Fig.~\ref{fig:file-structure}) (\textbf{R2}, \textbf{R9}). In addition, the control panel lets users visualize single or dual groups of selected ensembles and choose chart configurations, including parameter(s) to visualize, limb side, and gait cycle (\textbf{R9}) (Fig.~\ref{fig:teaser}-E). All visualizations in the different views are linked via cross-view interactions that allow highlighting of trials for different gait variables for more in-depth analysis (\textbf{R9}). We provide design justifications for the three views and cross-view interactions. In addition, we report all the alternative design choices in Fig. 2 in the appendix section.
The three views of the visualization frontend enable users to compare between two groups of ensembles (\textbf{R5}), analyze disease progression (\textbf{R6}), explore statistical measures (\textbf{R7}), and find and highlight anomalies (\textbf{R8}).

 \begin{figure}[t]
  \centering
  \includegraphics[width=0.7\linewidth]{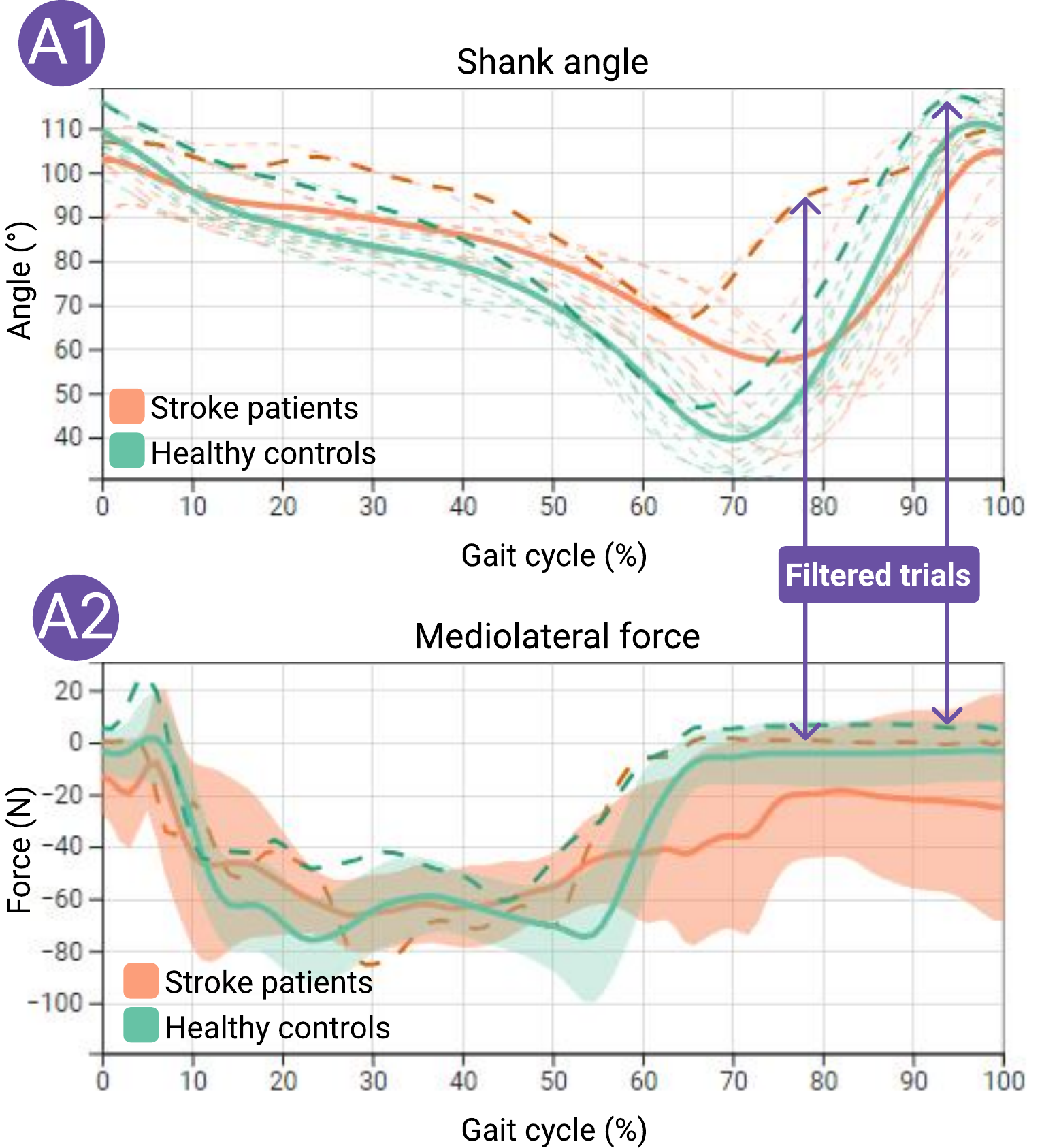}
  \caption{Time-series ensemble view. Displays the ensemble mean (green and orange solid lines) along with - \lower0.2em\hbox{\includegraphics[width=1em]{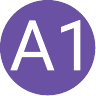}} individual trials' data (faded dashed lines) or \lower0.2em\hbox{\includegraphics[width=1em]{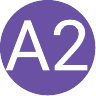}} the confidence interval - for two different groups (i.e., stroke patients and healthy controls). Users can filter and highlight a trial in \lower0.2em\hbox{\includegraphics[width=1em]{icons/a1.pdf}} by clicking on a faded dashed line, which will then appear as a bold dashed line within its view, other time-series ensemble views (e.g., \lower0.2em\hbox{\includegraphics[width=1em]{icons/a2.pdf}}), and the spatiotemporal summary view (Fig. \ref{fig:spatiotemporal-summary-view}). The filtered trials are also highlighted in the spatiotemporal distribution view (Fig. \ref{fig:spatiotemporal-distribution-view}) with rectangular brushes, indicating their range of values for each single-valued spatiotemporal parameter. Conversely, adjusting the rectangular brushing on any box plot in the spatiotemporal distribution view will highlight a different set of trials in the time series ensemble views (\lower0.2em\hbox{\includegraphics[width=1em]{icons/a1.pdf}}, \lower0.2em\hbox{\includegraphics[width=1em]{icons/a2.pdf}}) and spatiotemporal summary view (Fig. \ref{fig:spatiotemporal-summary-view}).}

  \label{fig:time-series-ensemble-view}
\end{figure}

\subsubsection{Time-series ensemble view}
\label{fig:subsubsec-time-series-ensemble-view}
The frontend contains four containers for time-series ensemble views (Fig.~\ref{fig:teaser} F1-F4), which display line charts for time-series ensemble data based on different parameter configurations selected in the control panel. Users can select any type of time-series data (e.g., motion data, joint angles, ground reaction forces)  to display in the time-series ensemble views (\textbf{R4}). The view allows the creation of two different ensemble groups and a comparison of their time-series characteristics (\textbf{R5}). Users can also select trials of the same patient from two different time periods to assess the progression of their time-series characteristics (\textbf{R6}). The view displays the ensemble mean of selected trials along with all individual trials (Fig.~\ref{fig:time-series-ensemble-view}-A1) or the confidence interval (Fig.~\ref{fig:time-series-ensemble-view}-A2) of the ensemble (\textbf{R7}). Hovering over a line highlights it and reveals the corresponding patient and trial id (Fig.~\ref{fig:teaser}-I1) with a tooltip, facilitating the identification of outliers and anomalies (\textbf{R8, R9}). In case of line with confidence intervals, hovering displays a tooltip showing the numeric values of the mean, lower bound, and upper bound at a specific time point for the selected gait parameter (Fig.~\ref{fig:teaser}-I2) (\textbf{R7, R9}).

\emph{Justification.}
Initially, we considered displaying multiple variables within a single chart (Fig. 2-A1 in the appendix) and using side-by-side plots for group comparisons (Fig. 2-A2 in the appendix). However, experts found it difficult to distinguish multiple variables (e.g., vertical force, mediolateral force) within one chart, and comparison using side-by-side plots was challenging for them. As a result, we transitioned to four separate time-series ensemble views (Fig.~\ref{fig:teaser} F1–F4), each dedicated to a single variable. Each view allows comparisons between up to two groups of trials (e.g., stroke, healthy).
Additionally, our initial approach displayed only the ensemble mean of a variable (Fig. 2-B1 in the appendix). However, experts preferred to see all individual trials to better identify anomalies, leading us to incorporate both individual trials and the ensemble mean (Fig. \ref{fig:time-series-ensemble-view}-A1). To further facilitate finding anomalies, we added a hover functionality that highlights a trial and displays its corresponding patient and trial ID (Fig.~\ref{fig:teaser}-I1).
For tasks not focused on finding anomalies but rather on understanding distribution patterns, experts requested a less cluttered view. In response, we introduced a mode displaying the mean with a confidence interval (Fig. \ref{fig:time-series-ensemble-view}-A2) and a tooltip that provides numerical values for the mean and confidence bounds (Fig.~\ref{fig:teaser}-I2).






\subsubsection{Spatiotemporal summary view} 
\begin{figure}[t]
  \centering
  \includegraphics[width=0.7\linewidth]{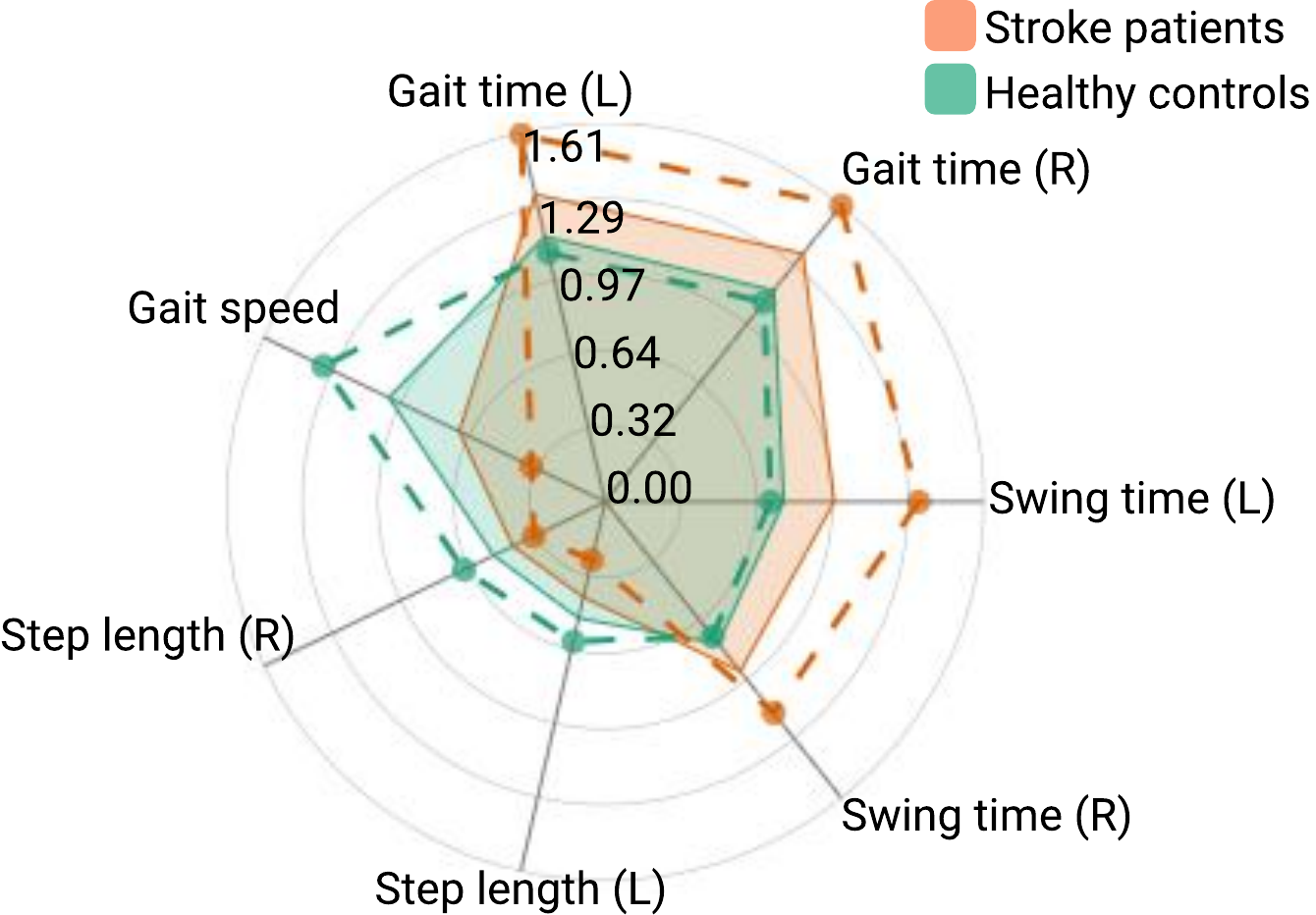}
  \caption{Spatiotemporal summary view. Displays the ensemble mean (opaque green and orange areas) of spatiotemporal parameters for selected lists of trial data for two different groups. When users filter trials by clicking on faded dashed lines from the time series ensemble view (Fig. \ref{fig:time-series-ensemble-view}) or brushing the range of a spatiotemporal parameter from one of the box plots in the spatiotemporal distribution view (Fig. \ref{fig:spatiotemporal-distribution-view}), the filtered trials get highlighted with bold dashed lines in this view.}
  \label{fig:spatiotemporal-summary-view}
\end{figure}

The ensemble mean of the spatiotemporal gait parameters is displayed using a radar chart in this view (Fig.~\ref{fig:spatiotemporal-summary-view}). This view allows users to effectively analyze multivariate spatiotemporal gait parameters (\textbf{R4}) and compare them between two groups (\textbf{R5}). Additionally, users can select data from the same patient across two time periods to examine the progression in spatiotemporal parameters over time (\textbf{R6}). Hovering over a parameter in the radar chart displays a tooltip with the numeric mean of the two ensembles for that parameter (Fig.~\ref{fig:teaser}-I3) (\textbf{R9}).

\begin{figure}[b]
  \centering
  \includegraphics[width=0.9\linewidth]{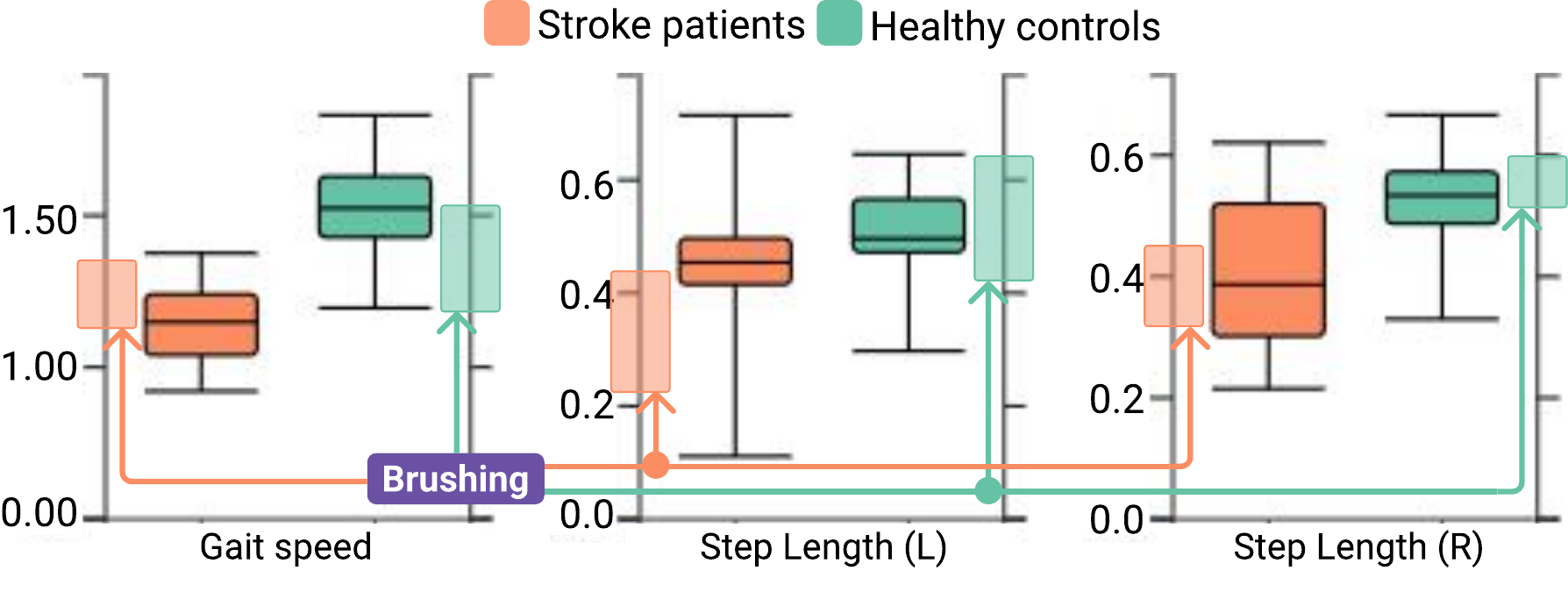}
  \caption{Spatiotemporal distribution view. This view features dual box plots for each spatiotemporal parameter, displaying the distribution for an ensemble of selected trials across two groups. Users can filter each plot for either group using rectangular brushes. Filtering one box plot will subsequently update the other box plots, as well as the time-series ensemble view (Fig.~\ref{fig:time-series-ensemble-view}) and the spatiotemporal summary view (Fig.~\ref{fig:spatiotemporal-summary-view}).}
  \label{fig:spatiotemporal-distribution-view}
\end{figure}

\emph{Justification.} 
The experts supported the survey results favoring a radar chart, noting that it provides a quick overview of how two ensembles differ across spatiotemporal parameters. While box plots (Fig.~\ref{fig:spatiotemporal-distribution-view}) show distribution differences, the radar chart (Fig.~\ref{fig:spatiotemporal-summary-view}) enables easier comparison of multiple variables at one glance. Experts also found it more intuitive than box plots for assessing disease progression for spatiotemporal parameters. Initially, we considered displaying more than two ensembles in the radar chart (Fig. 2-C1 in the appendix), but the experts noted that too many colors made the chart difficult to interpret and mentioned that comparing two ensembles was sufficient for their purposes. We also added the hover functionality (Fig.~\ref{fig:teaser}-I3), as the experts wanted to see exact numeric values on demand.


\subsubsection{Spatiotemporal distribution view} The distribution of multivariate spatiotemporal parameters (\textbf{R4}), including min, max, median, upper, and lower percentiles (\textbf{R7}), for the selected ensembles of trials is displayed using dual box plots in this view (Fig.~\ref{fig:spatiotemporal-distribution-view}). Users can also compare the distribution of two patient groups using this view (\textbf{R5}). Hovering over any box displays a tooltip with the statistical distribution of the selected spatiotemporal parameter (Fig.~\ref{fig:teaser}-I4) (\textbf{R7, R9}).



\emph{Justification.} The experts emphasized that box plots are essential for detailed statistical analysis of spatiotemporal parameters and are the most familiar visualization among gait practitioners. Initially, we displayed a single box plot per parameter (Fig. 2-E1 in the appendix), but the experts requested the ability to compare distributions between two groups, leading to the adoption of dual box plots (Fig.~\ref{fig:spatiotemporal-distribution-view}). To meet their need for numeric values on demand, we also added a tooltip (Fig.~\ref{fig:teaser}-I4) that appears on hover.


\subsubsection{Cross-view interactions} The frontend of VIGMA features several cross-view interactions, enabling interactive analysis between multiple coordinated views~\cite{munzner2014visualization}.

\begin{figure}[t]
  \centering
  \includegraphics[width=\linewidth]{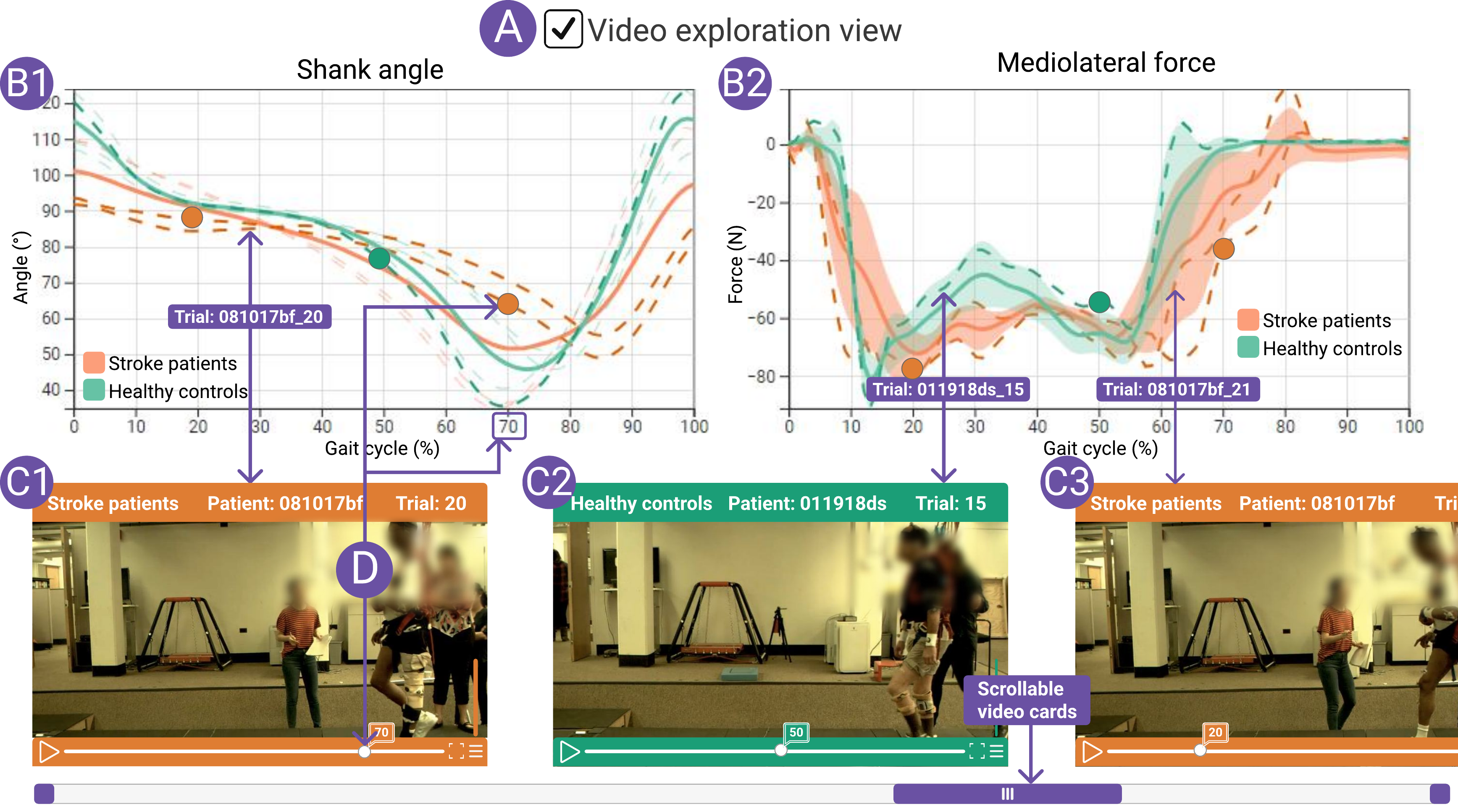}
  \caption{Access to raw trial videos \lower0.2em\hbox{\includegraphics[width=1em]{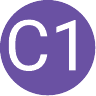}} - \lower0.2em\hbox{\includegraphics[width=1em]{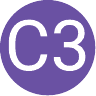}} is provided through a scrollable video exploration view. Users can toggle between spatiotemporal views (Fig.~\ref{fig:spatiotemporal-summary-view}, Fig.~\ref{fig:spatiotemporal-distribution-view}) and the video exploration view using a checkbox \lower0.2em\hbox{\includegraphics[width=1em]{icons/a.pdf}}. Videos appear in the video exploration view when users click on and highlight trials in the time-series ensemble views \lower0.2em\hbox{\includegraphics[width=1em]{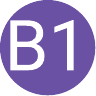}}, \lower0.2em\hbox{\includegraphics[width=1em]{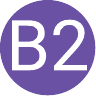}}. The video cards \lower0.2em\hbox{\includegraphics[width=1em]{icons/c1.pdf}} - \lower0.2em\hbox{\includegraphics[width=1em]{icons/c3.pdf}} are color-coded by patient group and display patient and trial id, and patient group information in the header. Videos are bi-directionally linked with the time-series ensemble views, with a circular marker \lower0.2em\hbox{\includegraphics[width=1em]{icons/d.pdf}} appearing in the time-series ensemble views to indicate the corresponding time in the video.}
  \label{fig:video-exploration-view}
\end{figure}

\emph{i) Clicking.} Users can highlight one or more trials by clicking on them in the time-series ensemble view, which renders them as bold dashed lines (Fig.~\ref{fig:time-series-ensemble-view}-A1, A2). The highlighted trials will also appear in bold dashed lines in all other time-series ensemble views and the spatiotemporal summary view (Fig.~\ref{fig:spatiotemporal-summary-view}). Additionally, the spatiotemporal distribution view will display a brush indicating the range of values for the highlighted trials (Fig.~\ref{fig:spatiotemporal-distribution-view}). Clicking on a trial in the time-series ensemble view also provides access to the raw video for that trial within the selected gait cycle timeframe (Fig.~\ref{fig:video-exploration-view}) (\textbf{R2}).

\emph{ii) Brushing.} The highlighting of trials across views works bi-directionally. If users brush a range in any box plot within the spatiotemporal distribution view (Fig.~\ref{fig:spatiotemporal-distribution-view}), the corresponding trials within the brush range will be highlighted in all time-series ensemble views (Fig.~\ref{fig:time-series-ensemble-view}). Additionally, the highlighted ensemble mean in the spatiotemporal summary view will update accordingly, displayed as bold dashed lines (Fig.~\ref{fig:spatiotemporal-summary-view}).

\emph{iii) Animated transition.} Users can play or pause raw videos from the video exploration view, which triggers a circular marker to appear on the time-series ensemble views for the corresponding trial (Fig.~\ref{fig:video-exploration-view}). As the video plays,  a circular marker dynamically moves across all time-series ensemble views corresponding to that trial.

%

\emph{Justification.} The experts wanted to analyze correlations between multivariate gait parameters across different views to gain deeper insights into tasks such as disease progression, outlier detection, and group comparisons. 
To support this, we implemented a click interaction within the time-series ensemble view. While hovering provided only temporary focus, clicking allowed users to highlight selected trials persistently across all views (Fig.~\ref{fig:time-series-ensemble-view}-A1, A2). For analyzing the distribution of spatiotemporal parameters and exploring their behavior across other views, we incorporated a brushing interaction in the dual box plots (Fig.~\ref{fig:spatiotemporal-distribution-view}). To provide access to raw videos of trials, we initially considered using pop-ups (Fig. 2-F1 in the appendix) linked to clicked lines in the time-series ensemble views. However, the experts found pop-ups distracting because they required switching focus to a separate window, breaking the connection between the video and the corresponding visualizations. Since the videos are tied only to the time-series parameters, instead of popups, we designed a video exploration view (Fig.~\ref{fig:video-exploration-view}) that displays videos in a scrollable format in lieu of the spatiotemporal summary and distribution views. Lastly, the experts wanted to maintain context between the video and the corresponding time-series data. We introduced circular markers (Fig.~\ref{fig:video-exploration-view}-D) that move along the time-series ensemble views, synchronized with the playback of the trial video, allowing users to visualize dynamic changes in time-series parameters as the video progresses.

\subsection{Implementation}
\label{subsec:implementation}
The system was implemented using Flask for the backend and React with D3.js~\cite{bostock2011d3} for the frontend. The computational notebooks employ a custom Python library we developed, which utilizes numpy, Pandas, SciPy, C3D, Scikit-learn, and plotly.
\section{Evaluation}
\label{sec:evaluation}

We evaluated the VIGMA framework through three usage scenarios created in collaboration with domain experts and experts' quantitative and qualitative feedback on the framework's understandability and usefulness.


\subsection{Usage Scenarios}
\label{subsec:usage}
We designed the usage scenarios in collaboration with two gait researchers (co-authors of the paper), using the necessary data from their research lab. The lab focuses on fall prevention, motor control, and stroke rehabilitation, collecting multivariate gait data from a diverse range of patients, including children, young adults, healthy older adults, and stroke patients. 
Through in-person meetings with the researchers, we identified common practices performed by gait practitioners across various labs and clinics, including:

\begin{itemize}[noitemsep, nolistsep]
    \item Addressing data quality issues due to hardware limitations, necessitating the identification and correction of such errors.
    \item Analyzing data from patients with disorders (e.g., stroke, parkinson) at intervals (e.g., 6 or 12 months) to identify rehabilitation progress.
    \item Comparing data from patients with disorders to data from healthy subjects to identify deviations.
\end{itemize}

Considering these common practices, we collected data from twenty subjects, among which ten were healthy subjects (aged from 61 to 76 years, mean $65.50 \pm 3.91$; body mass $84.72 \pm 14.49$ kg; height $1.68 \pm 0.08$ m; 6 males; 4 females) and the other ten were stroke patients (aged from 52 to 69 years, mean $61.20 \pm 4.73$; body mass $85.24 \pm 18.40$ kg; height $1.72 \pm 0.09$ m; 7 males; 3 females). The data of stroke patients was collected at the time of diagnosis and again after a 6-month interval. 
For the scenarios, we considered the following types of data: 3D motion, joint angles, ground reaction forces, and spatiotemporal parameters. 3D motion data and ground reaction forces were collected using motion capture cameras and force plates during regular walking trials  
on a 7-meter walkway, as illustrated in Fig.~\ref{fig:background}. The collected data, originally in TRC, MAT, and C3D formats, were standardized and converted to CSV using our framework. In addition, we utilized computational notebooks to extract joint angles from motion data, step times from ground reaction forces, and spatiotemporal parameters from joint angles and step times.

The collected data was then used in the following usage scenarios: (1) identifying and correcting erroneous data, (2) stroke rehabilitation analysis, and (3) comparing healthy subjects and stroke patients.
Stroke patients and healthy subjects, whose data were collected, were informed about the purposes of the study and provided written informed consent. The study was approved by UIC's IRB. Data collected for healthy controls and stroke patients fall under IRB protocol \#2016-0887 and \#2016-0933, respectively.

\begin{figure}[t]
  \centering
  \includegraphics[width=1\linewidth]{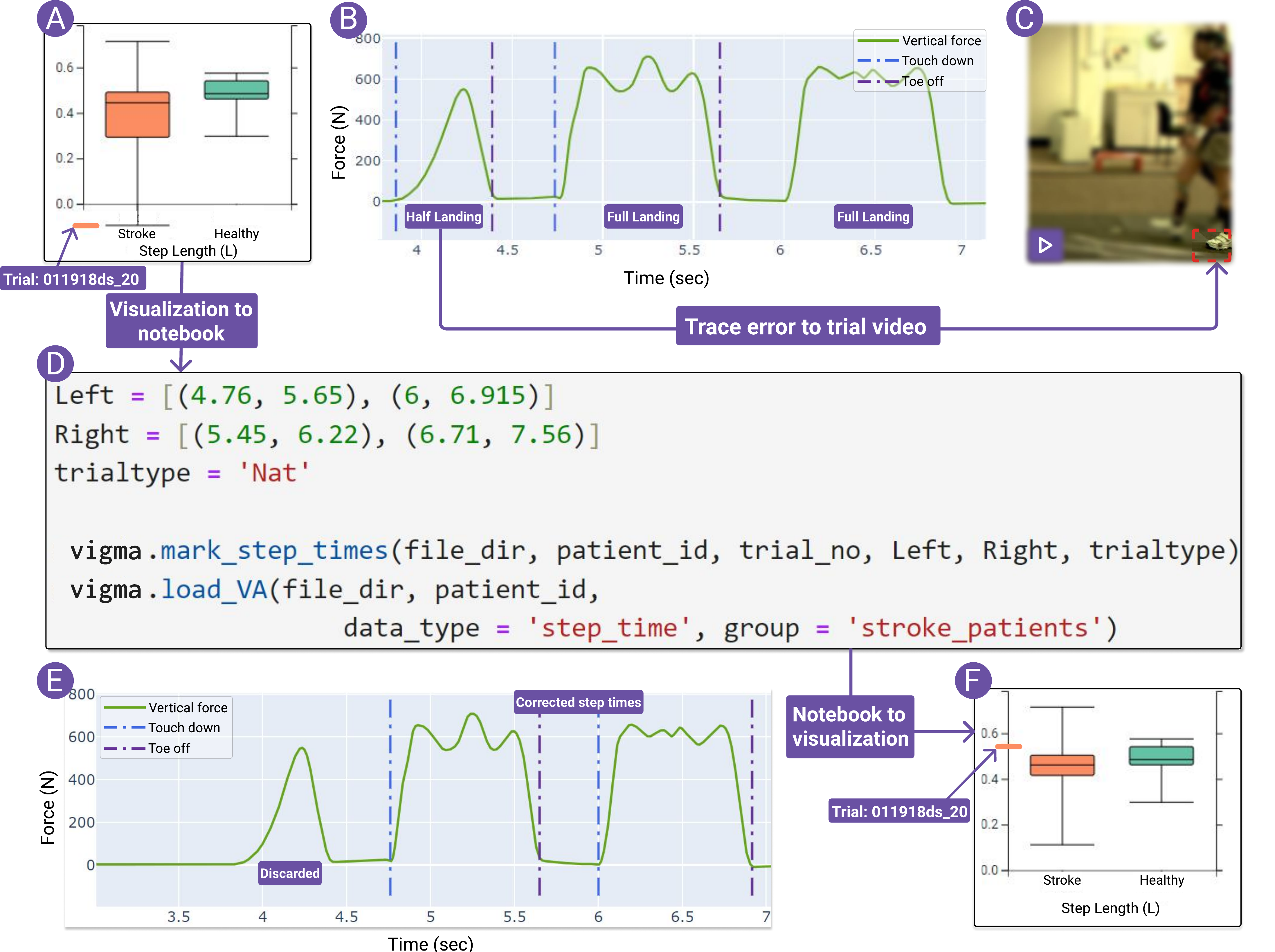}
  \caption{Usage scenario 1. \lower0.2em\hbox{\includegraphics[width=1em]{icons/a.pdf}} User identifies an error of negative step length in one of the stroke patient's trials from the spatiotemporal distribution view. \lower0.2em\hbox{\includegraphics[width=1em]{icons/b.pdf}} By examining vertical force and step times (touchdowns and toe-offs), they determine the cause of the error to be half landing on the platform and \lower0.2em\hbox{\includegraphics[width=1em]{icons/c.pdf}} validate it against the trial video using video exploration view. Afterwards, \lower0.2em\hbox{\includegraphics[width=1em]{icons/d.pdf}} user corrects the error by using the VIGMA library in a notebook,  \lower0.2em\hbox{\includegraphics[width=1em]{icons/e.pdf}} checks if the half landing is properly discarded and \lower0.2em\hbox{\includegraphics[width=1em]{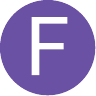}} returns to the visualization to see the updated value.}
  \label{fig:case-study-1}
\end{figure}

\subsubsection{Usage scenario 1: Identifying and correcting erroneous data}
\label{subsubsec:usage_1}
Anomalies, such as outliers and errors, are common occurrences in the process of collecting gait data. For the first usage scenario, we began by selecting ten stroke patients and ten healthy patients as two separate groups using the control panel and visualizing the spatiotemporal parameters using the spatiotemporal distribution view. We observed an instance of negative step length (left foot) for one of the trials from the stroke patients and identified the erroneous patient and trial id by brushing the box plot (Fig.~\ref{fig:case-study-1}-A) (\textbf{R8}). Analyzing the vertical force for that trial, we identified a half landing at the beginning of the trial (Fig.~\ref{fig:case-study-1}-B) and validated our speculation by tracing it to the original trial video using the video exploration view (Fig.~\ref{fig:case-study-1}-C). To correct the issue, we used our Python API in a notebook to mark step times by discarding the half landing and loaded the corrected data to the visual analytics system (Fig.~\ref{fig:case-study-1}-D, E) (\textbf{R1}). Returning to the spatiotemporal distribution view, we then identified the correct step length for the left foot for that trial (\textbf{R8}).

This usage scenario demonstrates how users can leverage notebooks and the visualization interface to seamlessly navigate between data processing and visualization tasks.

\subsubsection{Usage scenario 2: Stroke rehabilitation analysis} 
The second usage scenario demonstrates the system's capability to analyze stroke rehabilitation progress over time, a key focus in motion and gait studies (\textbf{R6}).

\begin{figure}[t]
  \centering
  \includegraphics[width=1\linewidth]{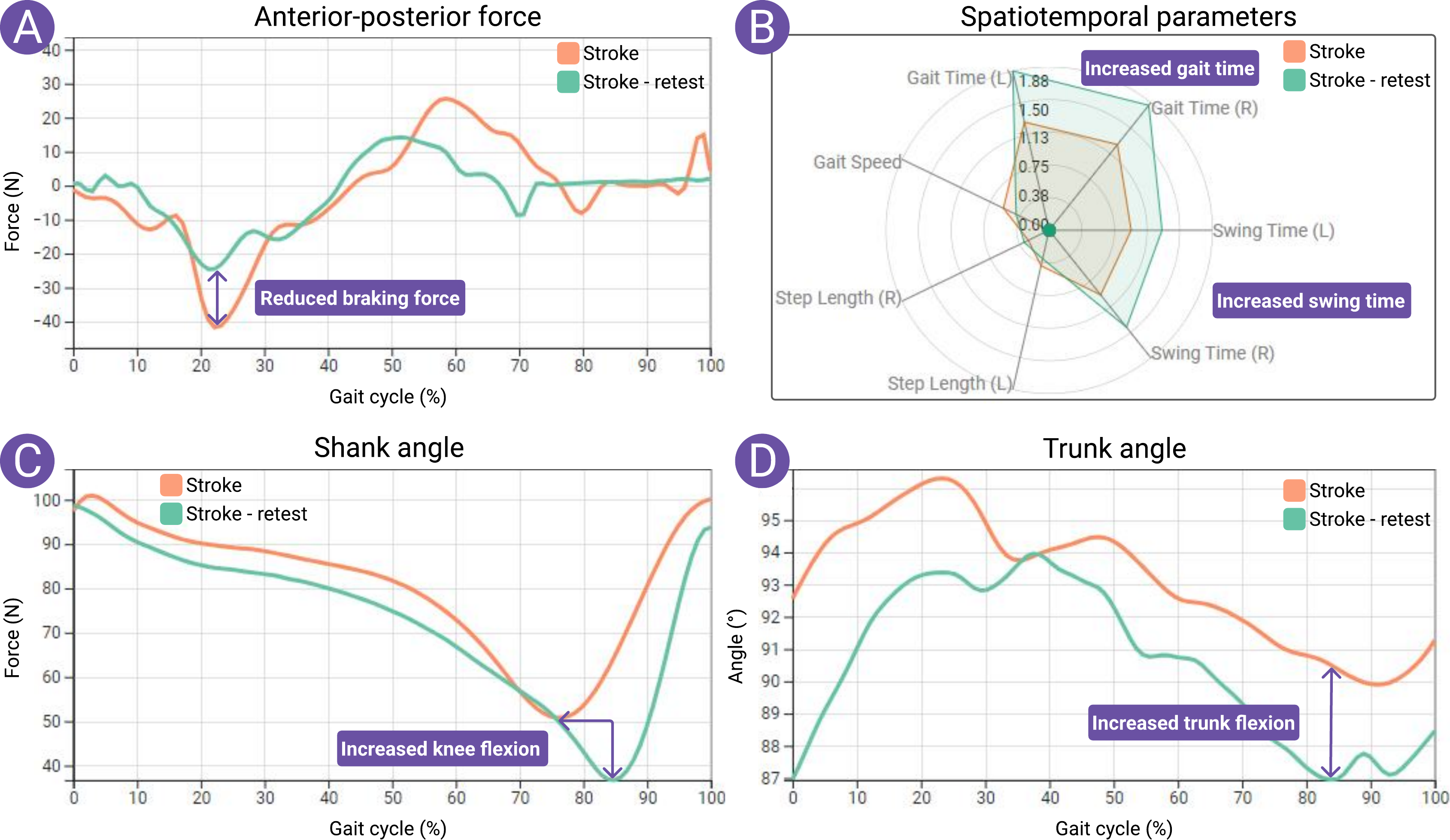}
  \caption{Usage scenario 2. Identifying the rehabilitation progress of a stroke patient over time. Orange represents data at the time of the stroke, while green represents the same patient's data after 6 months. The patient received instructions to increase gait time and reduce braking force to minimize fall risk. Improvements in the post-6-month data show \lower0.2em\hbox{\includegraphics[width=1em]{icons/a.pdf}} reduced braking force, \lower0.2em\hbox{\includegraphics[width=1em]{icons/b.pdf}} increased gait and swing time, and reduced gait speed. Additionally, the patient exhibited more \lower0.2em\hbox{\includegraphics[width=1em]{icons/c.pdf}} knee and \lower0.2em\hbox{\includegraphics[width=1em]{icons/d.pdf}} trunk flexion, indicating improved balance and stability in their gait.}
  \label{fig:case-study-2}
\end{figure}

First, we selected the data using the control panel for one stroke patient at two different time points: the first trial and after six months. The patient was instructed to reduce braking force and gait speed to minimize fall risk during walking. We generated three time-series ensemble views for anterior-posterior force (Fig.~\ref{fig:case-study-2}-A), shank angle (Fig.~\ref{fig:case-study-2}-C), and trunk angle (Fig.~\ref{fig:case-study-2}-D), as well as a spatiotemporal-summary view (Fig.~\ref{fig:case-study-2}-B) to compare the spatiotemporal parameters over the six-month period. Analyzing the views, we observed a reduction in the patient's braking force, particularly during the initial phase of the gait cycle (Fig.~\ref{fig:case-study-2}-A). Additionally, the overall gait time and swing time for both feet increased after six months, resulting in a reduced gait speed. This indicated an overall improvement in the patient’s gait, reducing the risk of falls. Furthermore, analyzing the shank and trunk angles, we observed an increase in both knee flexion (Fig.~\ref{fig:case-study-2}-C) and trunk flexion (Fig.~\ref{fig:case-study-2}-D), indicating improved flexibility and stability in these limbs.

\subsubsection{Usage scenario 3: Comparing healthy subjects and stroke patients} Practitioners frequently compare gait characteristics across different groups of patients to identify key differences and to analyze how an individual patient's data aligns or differs with that of a specific patient group. Our third usage scenario specifically examines this aspect in detail (\textbf{R5}).

\begin{figure}[t]
  \centering
  \includegraphics[width=1\linewidth]{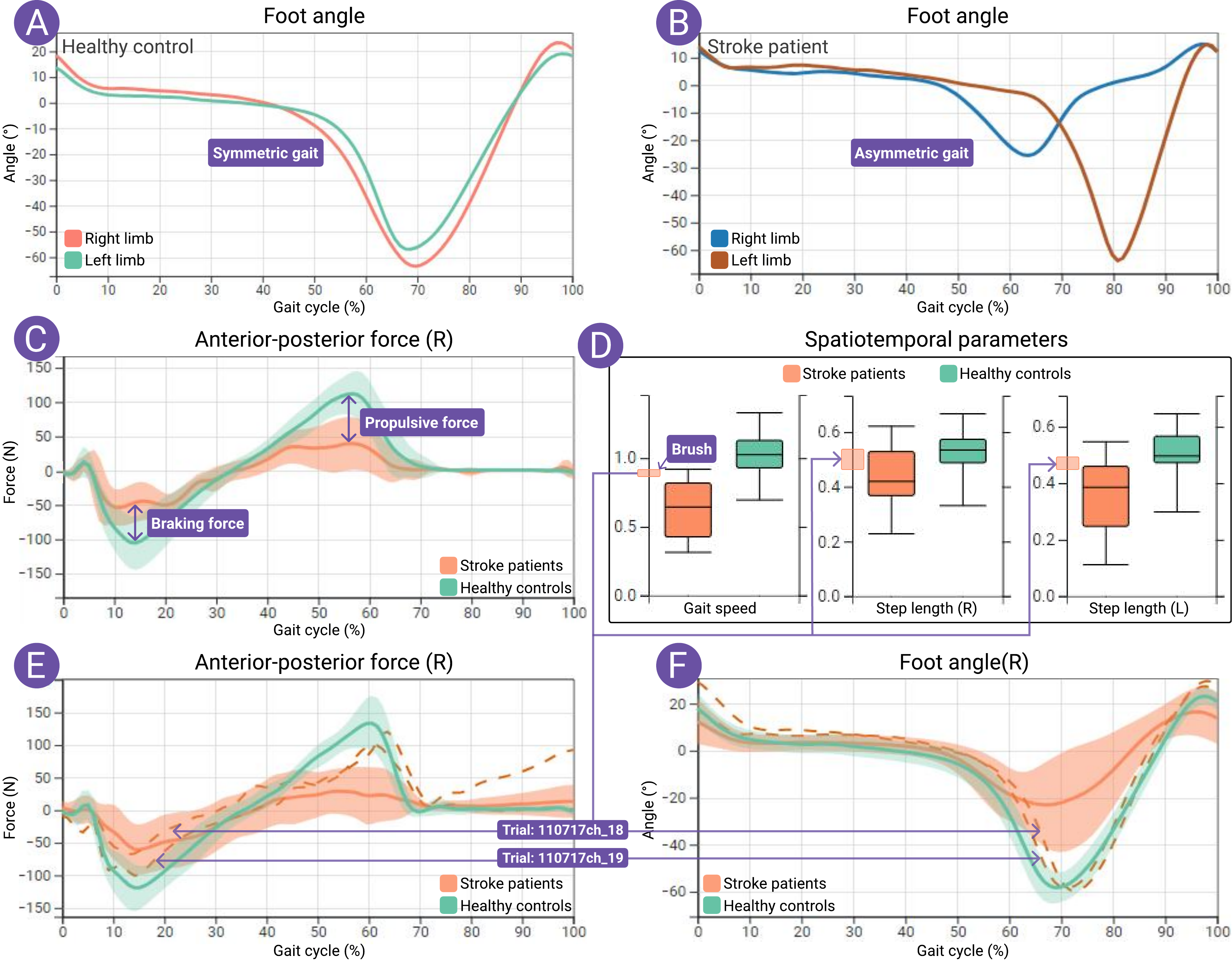}
  \caption{Usage scenario 3. Comparing the foot angle of \lower0.2em\hbox{\includegraphics[width=1em]{icons/a.pdf}} a healthy older adult with \lower0.2em\hbox{\includegraphics[width=1em]{icons/b.pdf}} a right-sided paretic stroke patient to identify the magnitude of asymmetry in the stroke patient's gait. The user then ensembles trials from five right-sided paretic stroke patients and five healthy controls to compare their \lower0.2em\hbox{\includegraphics[width=1em]{icons/c.pdf}} right-sided anterior-posterior force and \lower0.2em\hbox{\includegraphics[width=1em]{icons/d.pdf}} spatiotemporal parameters. Additionally, the user brushes the box plot for stroke patients with higher gait speeds and highlights the corresponding trials for \lower0.2em\hbox{\includegraphics[width=1em]{icons/e.pdf}} right-sided anterior-posterior force and \lower0.2em\hbox{\includegraphics[width=1em]{icons/f.pdf}} right-sided foot angle. The user finds that the highlighted trials belong to the same stroke patient whose gait is more similar to those of healthy controls, suggesting they require a less intensive rehabilitation strategy.}
  \label{fig:case-study-3}
\end{figure}

We began by comparing the left and right foot angles in a right-sided paretic stroke patient (Fig.~\ref{fig:case-study-3}-B) to a healthy control (Fig.~\ref{fig:case-study-3}-A). We observed that the healthy control shows consistent patterns for both the left and right foot, suggesting a balanced and symmetrical gait. However, the stroke patient exhibited clear asymmetry between their paretic and non-paretic sides, and we were able to analyze the degree of asymmetry in the foot angles.

For the next part, we merged data from five right-sided paretic stroke patients and compared their right anterior-posterior force (Fig.~\ref{fig:case-study-3}-C) and spatiotemporal parameter distribution (Fig.~\ref{fig:case-study-3}-D) with those of five healthy controls. We observed that stroke patients exhibited a flatter anterior-posterior force curve with less pronounced peak (propulsive force) and trough (braking force) than healthy controls. The lower propulsive force in stroke patients suggests that they have difficulty in generating forward momentum during walking, which impacts their gait speed. The spatiotemporal distribution view supports this hypothesis, revealing that the gait speed distribution for stroke patients is significantly lower than that of healthy controls.

Lastly, we analyzed the outliers in stroke trials that exhibited higher gait speeds by examining the box plot of gait speed in Fig.~\ref{fig:case-study-3}-D (\textbf{R8}). We discovered that these trials had higher step lengths compared to most stroke patients by visualizing the distributions of left and right step lengths (Fig.~\ref{fig:case-study-3}-D). Additionally, we observed that their right anterior-posterior force (Fig.~\ref{fig:case-study-3}-E) and foot angles (Fig.~\ref{fig:case-study-3}-F) were more similar to those of healthy controls than to other stroke patients. By tracing these trials to their corresponding videos, we identified that they all belonged to the same patient. Further analysis of these trial videos revealed that this patient had already improved their gait characteristics to a certain extent and required less intensive rehabilitation strategies.

\subsection{Experts' Feedback}
\label{subsec:feedback}
After creating the usage scenarios, we engaged with five experts from different research labs and clinics to get their feedback on VIGMA. The five experts are not co-authors of the paper.
Table~\ref{tab:participant_info} summarizes their background.

Each expert was exposed to all three usage scenarios, with each session lasting approximately 1.5 hours to complete. To save time, three separate instances of VIGMA, preloaded with the required data for each usage scenario, were opened before starting the session. The experts were asked to run the usage scenarios themselves by executing the computational notebooks and interacting with the visualization frontend, receiving assistance if they encountered any issues. 
Two of them participated in the study in person, while the other three joined via Zoom meetings with desktop sharing. 
It is important to note that the experts who tested the usage scenarios and provided feedback were different from the two researchers who collaborated in the creation of the scenarios.

We gathered both qualitative and quantitative feedback from each expert. The sessions for testing the usage scenarios were open-ended, allowing the experts to provide qualitative feedback both during and after the sessions. At the end of the session, the experts were asked to complete an online questionnaire to provide quantitative feedback.

\subsubsection{Qualitative feedback} 
Overall, our framework received positive qualitative feedback from the experts. One of the senior clinicians and faculty members affirmed: ``\textit{As a clinician I feel the visualization tool is easy to understand and very useful to plot and visualize data. I would definitely use this to plot my information.}'' The clinician found the studies to be enlightening, particularly in demonstrating how the different views could help analyze the rehabilitation progress of stroke patients and design more effective rehabilitation strategies. Another senior clinician echoed this sentiment, stating, ``\textit{the tool comes with awesome flexibility. Very helpful in detailed analysis.}'' This clinician was particularly impressed by the flexibility of processing and analyzing various types of multivariate gait data using the system. 

Ongoing PhD students and researchers found the tool to be equally useful. One of them remarked ``\textit{the tool can serve in both clinical and research capacities.}'' Another student noted ``\textit{a system like this will be extremely useful for both clinicians and researchers. From a clinician perspective, it will expand our ability to perform detailed gait analyses without requiring an expert biomechanist or computer engineer - expanding also the number of people who can be positively impacted by this system. Additionally, as a PhD Candidate, I think that it will help people like me work smarter and expand our ability to analyze more data and obtain results which are more impactful.}'' We received highly positive remarks for the inclusion of videos in the system. One researcher commented that ``\textit{having access to videos in the system is very helpful in detecting and validating gait abnormalities or deviations.}''

We received mixed feedback regarding the computational notebooks used for data processing tasks, with one of the experts mentioning, ``\textit{although I have very little experience in coding, the notebook side of things seemed very straightforward and easy to me.}'' On the other hand, one of the clinicians stated ``\textit{down the road, I think it would be useful to have buttons to represent the notebook functions for individuals with less computer science/engineering background.}'' Another clinician emphasized the need for proper instructions to effectively use the API, stating, ``\textit{this would require training or step-by-step instructions to ensure understanding.}''



\begin{table}[t]
\centering
\caption{Participants involved in the evaluation. Each expert tested all three usage scenarios and later provided feedback. The experts are not co-authors of the paper. Background abbreviations: F - Faculty, C - Clinician, R - Researcher, S - Student.}
\begin{tabular}{|l|c|c|l|}
\hline
\multicolumn{1}{|c|}{Background} & \multicolumn{1}{c|}{\begin{tabular}[c]{@{}c@{}}Years in \\ field\end{tabular}} & \multicolumn{1}{c|}{\begin{tabular}[c]{@{}c@{}}Knowledge in\\ gait (1-5)\end{tabular}} & \multicolumn{1}{c|}{\begin{tabular}[c]{@{}c@{}}Education level\end{tabular}} \\
\hline
F, C, R & 5 & 4 & PhD \\

S & 2 & 3 & PhD (Pursuing) \\

F, C & 10 & 4 & PhD \\

C, R & 1 & 2 & Masters \\

C, S & 6 & 4 & PhD (Pursuing) \\
\hline
\end{tabular}
\label{tab:participant_info}
\end{table}

\subsubsection{Quantitative feedback}
\begin{figure}[t]
  \centering
  \includegraphics[width=0.75\linewidth]{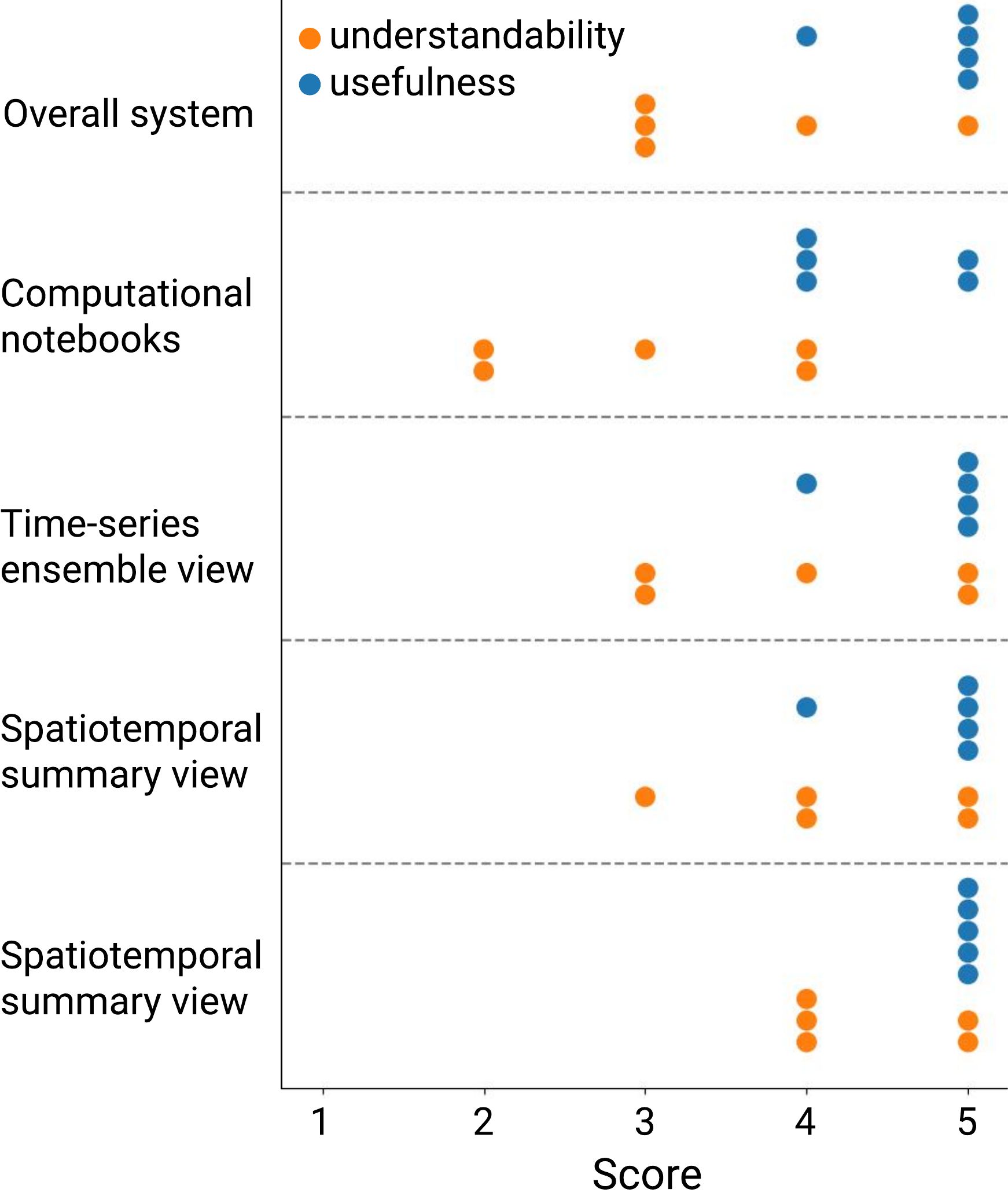}
  \caption{VIGMA's usefulness and understandability scores from experts on a Likert scale from 1 to 5.}
  \label{fig:likert-feedback}
\end{figure}

In addition to qualitative feedback, we asked the experts to complete an online questionnaire rating the usefulness and understandability of the whole framework and individual components on a Likert scale. The results are shown in Fig.~\ref{fig:likert-feedback}. 

Feedback was positive, especially regarding usefulness, with all ratings falling between 4 and 5. Understandability ratings varied more, with computational notebooks receiving lower scores compared to other components. This aligns with qualitative feedback from some experts, who noted a desire for the computational notebooks to be more simplified or to include proper documentation with examples for easier use of the API. Based on feedback from the experts, we have made the complete API documentation available on our GitHub page, including examples of one-line function calls to facilitate easy replication of data processing tasks.

\subsection{Comparison With the State of the Art}
\label{subsec:state-of-the-art}
We also compared VIGMA with existing tools across data processing, analysis, and visualization features. (see Table~\ref{tab:tools-and-techniques}). Here, we discuss tools selected from the table based on whether they partially or fully support at least 35\% of the features.
The tools include: (1) bioMechZoo~\cite{dixon2017biomechzoo}, a MATLAB toolbox for processing, analyzing, and visualizing multivariate gait data; (2) Vicon Nexus~\cite{vicon}, a multivariate gait data collection tool that utilizes markers, motion cameras, and force plates, providing some data processing and analysis capabilities; (3) TEMPLO~\cite{templo}, a markerless gait data collection tool allowing for single-trial analysis of different gait data types; (4) Visual3D~\cite{visual3d}, an all-in-one data analysis tool for processing, analyzing, and visualizing various types of multivariate gait data; (5) KAVAGait~\cite{wagner2018kavagait}, a gait analysis tool for storing and analyzing implicit knowledge of gait data such as ground reaction forces and spatiotemporal parameters through visualizations; and (6) Kinovea~\cite{kinovea}, a video analytics tool for gait analysis, primarily designed for sports.


Regarding data processing requirements, only Visual3D partially or fully meets most of the criteria. It offers limited format harmonization (C3D, ASCII), and while it supports feature extraction, noise filtering, and gait cycle normalization, it does not support missing value imputation (\textbf{R1}). Additionally, the data processing steps are performed through a complex interface and do not support the integration of computational notebooks based on Python or MATLAB (\textbf{R3}). Among the other tools, only bioMechZoo offers comprehensive support for feature extraction, missing value imputation, noise filtering, and gait cycle normalization (\textbf{R1}), and its MATLAB-based code can be integrated with notebooks (\textbf{R3}). However, it only supports MAT files for data formats (\textbf{R1}) and lacks organized access to patient data through a visual interface (\textbf{R2}). Additionally, both of these systems have lengthy documentation and complex code syntax for processing data. Finally, among the six selected tools, only TEMPLO and Kinovea provide access to raw video data (\textbf{R2}).

Concerning data analysis and visualization requirements, Visual3D supports statistical (\textbf{R7}) and anomaly analysis (\textbf{R8}) through non-interactive charts. However, limited interactions (\textbf{R9}) makes it inadequate for performing patient group comparisons (\textbf{R5}) or disease progression (\textbf{R6}). bioMechZoo faces the same limitations. KAVAGait offers interactive analysis (\textbf{R9}), but is limited to analyzing only a specific subset of gait data types (\textbf{R4}). KAVAGait also supports patient group comparisons (\textbf{R5}), but the analysis is limited to spatiotemporal parameters (\textbf{R4}). Lastly, of the six tools considered, only Kinovea is open source, underscoring a significant gap in the availability of open-source tools to drive research and clinical gait analysis.

\section{Conclusion}
\label{sec:conclusion}

This paper introduces VIGMA, an open-access visual analytics framework designed for multivariate gait analysis that integrates computational notebooks to facilitate data processing tasks and a frontend for visualization and data analysis tasks. The Python API offers one-liner function calls to simplify complex data processing tasks, and then seamlessly integrates processed data into the visual frontend, enabling comprehensive analysis and visualization of multivariate gait data. 
%
The usage scenarios demonstrate how the system achieves its goal of streamlining the integration between data processing, and data analysis and visualization, while efficiently handling common gait analysis tasks such as tracking disease progression and performing group comparisons.
VIGMA's open-access framework aims to foster greater collaboration between research labs and clinics by offering a one-stop solution for gait workflow.

\noindent \textbf{Limitations.} 
While promising, the current state of VIGMA presents some limitations. Some clinicians who participated in the usage scenarios highlighted that, although they found the one-liner function calls useful for data processing tasks, they would prefer something even simpler, and added the need for detailed documentation. To address this, we have made the documentation available on GitHub, including examples for easy replication. Additionally, the current usage scenarios are limited to data collected from one research lab and focus exclusively on stroke patients.

\noindent \textbf{Future work.} 
To address the complexity in computational notebooks highlighted in the experts' feedback, in addition to the documentation, following recent works~\cite{kery2020mage, choi2023towards}, we plan to integrate widgets and buttons into the computational notebooks to further simplify data processing tasks. Furthermore, we plan to create additional training materials, such as video tutorials, to make the framework more easily accessible to researchers and clinicians. We also plan to gather data from various labs and clinics, involving a more diverse group of patients (e.g., cerebral palsy patients, athletes), and include new usage scenarios that demonstrate the framework's versatility and effectiveness.
Also, as future work, we aim to explore new methodologies for developing open-access frameworks designed for widespread adoption, motivated by our prior efforts (e.g.,~\cite{moreira2024utk, moreira2025curio}), moving beyond the closed-source bespoke systems typically seen in design studies. While our current work takes a step in this direction by incorporating surveys and interviews with a broad range of experts, we aim to formalize this approach as a distinct methodology in the future.

\section*{Acknowledgments}

We thank the reviewers for their constructive comments and helpful suggestions.
Our work is supported by the National Institutes of Health (Award R01-HD088543, Administrative Supplement) and the National Science Foundation (Awards \#2320261, \#2330565, \#2411223).


\bibliographystyle{IEEEtran}
\bibliography{references}

\begin{IEEEbiographynophoto}{Kazi Shahrukh Omar} is Ph.D. student in Computer Science at the University of Illinois Chicago. His research interests include visualization and visual analytics, urban computing, and health informatics.
\end{IEEEbiographynophoto}

\begin{IEEEbiographynophoto}{Shuaijie Wang} is a Research Associate Professor in the Cognitive, Motor and Balance Rehabilitation Laboratory at the University of Illinois Chicago. He earned his Ph.D. from the University of Chinese Academy of Sciences. His expertise lies in biomedical engineering, with a focus on gait analysis, balance control, fall prevention, and fall risk assessment. His research involves understanding the mechanisms of falls, developing training programs to reduce fall risk in older adults, and using machine learning to model and assess fall risk.
\end{IEEEbiographynophoto}

\begin{IEEEbiographynophoto}{Ridhuparan Kungumaraju} is a Master’s student in Computer Science at the University of Illinois Chicago. His research interests include visual data science and software development.
\end{IEEEbiographynophoto}

\begin{IEEEbiographynophoto}{Tanvi Bhatt} is a Professor in the Department of Applied Health Sciences at the University of Illinois Chicago. She is the Director of the Cognitive, Motor and Balance Rehabilitation Laboratory and Co-Director of the Clinical Gait and Motion Analysis Laboratory. She received her Ph.D. in Movement Sciences/Motor Control from the University of Illinois at Chicago. Her research focuses on adaptive perturbation training for fall prevention, investigating the neuromechanical basis of balance recovery from slips and trips, and designing interventions to reduce fall risk.
\end{IEEEbiographynophoto}

\begin{IEEEbiographynophoto}{Fabio Miranda} is an Assistant Professor in the Department of Computer Science at the University of Illinois Chicago. He received his Ph.D. in Computer Science from New York University. His research focuses on large-scale data analysis, data structures, and data visualization, and visual analytics.
\end{IEEEbiographynophoto}
\vfill

\end{document}